\documentclass[onecolumn,showpacs,preprintnumbers,amsmath,amssymb]{revtex4}
\usepackage{graphicx}
\usepackage{dcolumn}
\usepackage{bm}
\usepackage{epsfig}

\newcommand{\bi}{\begin{itemize}}
\newcommand{\ei}{\end{itemize}}
\newcommand{\be}{\begin{eqnarray}}
\newcommand{\ee}{\end{eqnarray}}
\newcommand{\beq}{\begin{equation}}
\newcommand{\eeq}{\end{equation}}
\newcommand{\beqn}{\begin{equation*}}
\newcommand{\eeqn}{\end{equation*}}
\newcommand{\bbmatrix}{\left( \begin{array}}
\newcommand{\eematrix}{\end{array} \right)}

\begin{document}

\title{Final-state effect on x-ray photoelectron spectrum of nominally  $d^1$ and $n$-doped $d^0$ transition metal oxides}

\author{Chungwei Lin, Agham Posadas, Tobias Hadamek and Alexander A. Demkov\footnote{demkov@physics.utexas.edu}}
\date{\today}
\affiliation{Department of Physics, The University of Texas at Austin}
%538W 120th St NY, NY 10027}

\begin{abstract}
We investigate the x-ray photoelectron spectroscopy (XPS) of nominally $d^1$ and $n$-doped $d^0$ transition metal oxides 
including NbO$_2$, SrVO$_3$, and LaTiO$_3$ (nominally $d^1$), as well as $n$-doped SrTiO$_3$ (nominally $d^0$). 
In the case of single phase $d^1$ oxides, we find that the XPS spectra 
(specifically photoelectrons from Nb $3d$, V $2p$, Ti $2p$ core levels) all display at least two, 
and sometimes three distinct components, which can be consistently identified as $d^0$, $d^1$, and $d^2$
oxidation states (with decreasing order in binding energy). 
Electron doping increases the $d^2$ component but decreases the $d^0$ component, 
whereas hole doping reverses this trend; a single $d^1$ peak is never observed, and the $d^0$ peak is always present 
even in phase-pure samples. In the case of $n$-doped SrTiO$_3$, the $d^1$ component appears 
as a weak shoulder with respect to the main $d^0$ peak. 
We argue that these multiple peaks should be understood as being due to the final-state effect 
and are intrinsic to the materials. Their presence does not necessarily imply the existence of 
spatially localized ions of different oxidation states nor of separate phases.
A simple model is provided to illustrate this interpretation, 
and several experiments are discussed accordingly. 
The key parameter to determine the relative importance between the initial-state and final-state effects is also pointed out. 
\end{abstract}

\maketitle

%%%%%%%%%%%%%%%%%%%%%%%%%%%%%%%%%%%%%%%%%%%%%%%%%%%%%%%%%%%%%%%%%%%%%%%%%%%%

\section{Introduction}

X-ray photoelectron spectroscopy (XPS) is a very common {\em in-situ} and {\em ex-situ} tool used in modern laboratories
to probe the stoichiometry of a given material, as well as the oxidation states and local chemical environment of a given element
\cite{XPS_general, XPS_general2, CoreLevel, Hufner}.
As the core levels of different chemical elements easily differ by tens to hundreds of electron-volts (eV), 
the peaks in the photoelectron distribution as a function of kinetic energy provide us with information 
on which chemical elements are present and, to a very good approximation, their relative abundance in a sample.
When focusing on the photoelectron signals coming from one particular core level of one particular element, 
the different local environments around the targeted ions  can
result in a multi-peak structure, typically within an energy range of about 10 eV, from which the oxidation states of the probed element
can be inferred \cite{PhysRevB.38.6084, Gonzalez-Elipe_1988_Si, Miller_2002_C}.
A more sophisticated aspect of XPS is the electron screening due to the created core hole  \cite{CoreLevel}:
once a photoelectron is generated, the sample is left with a core hole (positively charged) that modifies 
the potential of valence electrons. The response of valence electrons to the core hole 
is usually referred to as the final-state effect, in the sense that the observed 
spectrum does not really correspond to that of the neutral sample before being irradiated, 
but rather to the energy spectrum in the presence of a core hole.
The typical lifetime of a core hole is about $10^{-15}$ s \cite{CoreLevel}, which results in
an energy broadening of $\sim$0.1 eV. Accordingly, peak features that are larger than 0.1 eV in the core-hole spectrum can, 
in principle, be  observed  and resolved.

The final-state effect introduces even more features and complexities to the XPS spectrum, as 
electron correlation is essential to the process of core-hole screening.
For example, the XPS spectra of a metallic system typically has an asymmetric shape (orthogonality catastrophe) 
when taking the scattering of the  core-hole potential into account \cite{PhysRevLett.18.1049, Mahan, Doniach}.
In addition, if the targeted ion has degenerate localized orbitals (such as $3d$ or $4f$ orbitals) in a metallic phase,
a uniform system also displays multiple XPS peaks. To properly describe such systems theoretically, an Anderson impurity model 
including both localized correlated orbitals and uncorrelated bath orbitals 
is required \cite{Kotani_1974, PhysRevB.28.4315, CoreLevel}.
For the transition metal (TM) oxides, the valence states have to include both oxygen $2p$ and TM $d$ orbitals,
as their energy difference and their mutual hopping amplitude are comparable in energy.
Therefore, a minimal model for XPS spectra of transition metal oxides includes a TM-O$_6$ cluster \cite{PhysRevB.33.8060, PhysRevB.45.1612,Kotani_93, CoreLevel}.
Although complicated, once the XPS spectrum is properly interpreted, it provides a quite good estimate 
of material-specific parameters such as inter-site hopping amplitude $t$ and Hubbard on-site repulsion $U$.

%\cite{PhysRevLett.55.418, CoreLevel}.

In this paper, we reexamine the origin of the multi-peak structure in the XPS spectra of 
nominally $d^1$ transition metal oxides including NbO$_2$, SrVO$_3$ \cite{SVO}, and LaTiO$_3$,
as well as that of lightly $n$-doped $d^0$ SrTiO$_3$ (STO)
\cite{PhysRevB.83.035410,/content/aip/journal/apl/100/26/10.1063/1.4731642, /content/aip/journal/jap/116/4/10.1063/1.4891225},
In particular, we propose a cluster-bath model and argue that it is the final-state effect
rather than the presence of multiple oxidation states 
that accounts for 
the observed multi-peak XPS structure in these materials. Based on our interpretation,
the multiple XPS peaks are intrinsic to the materials, and do not necessarily imply 
the existence of spatially localized ions with different oxidation states or of separate phases.
The rest of the paper is organized as follows.
In Section II we give a brief overview of the XPS core level spectra of these four oxides. In particular,
we distinguish between the initial-state effect and final-state effect. 
In Section III we present our experimental results and point out their common features and their implications. 
In Section IV we provide a simple model to illustrate the final-state effect, 
which is crucial to reconciling the seemingly conflicting observations.
Several experimental results are discussed accordingly.
The key dimensionless parameter to determine the relative
importance between initial-state and final-state effects is identified.
A brief conclusion is given in Section V.
In the Appendices we provide the details of our calculations.

\section{Overview of XPS}

\begin{figure}[http]
\epsfig{file=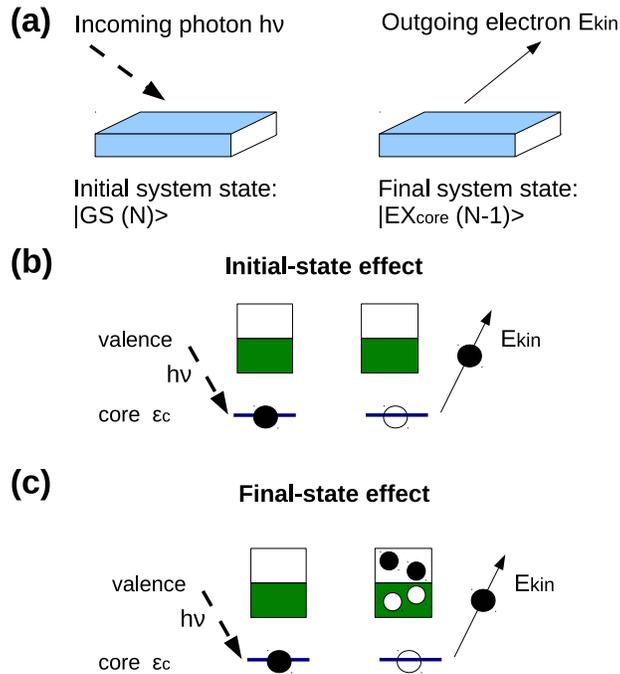, width = 0.5\textwidth}
\caption{ (Color Online) Illustration of XPS measurements. (a) (Left) The overall initial state contains an 
incoming photon of energy $h \nu$, and the system is in the ground state with the core level filled.
(Right) The overall final state contains an outgoing electron of kinetic energy $E_{kin}$, 
and the system is in some excited state with the core level empty.
(b) For the initial-state effect, the valence electrons are not affected by the created core hole, and therefore 
the binding energy is determined by the core-level energy $\epsilon_c$. 
(c) For the final-state effect, the valence electrons do respond to the created core hole. This 
can lead to multiple peaks around $\epsilon_c$ in the XPS spectrum. 
}
\label{fig:XPS_illu}
\end{figure}

In an XPS experiment, photons of energy $h \nu$ are directed to the sample and photoelectrons of 
kinetic energy $E_{kin}$ come out [see Fig.~\ref{fig:XPS_illu}(a)]. Energy conservation requires that
\beq
h \nu + E_{GS} (N) = E_{kin} + E_{core} (N-1) + \phi.
\label{eqn:XPS_basic}
\eeq
Here $E_{GS} (N)$ is the ground state energy of the sample with the filled core level, 
$E_{core} (N-1)$ is the energy with a core hole ($N-1$ is used to denote the presence of a core hole), and $\phi$ is the work function.
%$N$ and $N-1$ are respectively the numbers of electrons in the sample without and with the core hole.
By shifting the kinetic energy by $E_{kin} \rightarrow \omega = E_{kin} + \phi - h\nu$,
the photoelectron intensity as a function of $\omega$  is given by
\beq 
\begin{split}
\rho (\omega) &= \sum_n |\langle n (N-1) | c |GS \rangle|^2 \times \delta(\omega - [E_{GS} (N) - E_{core,n} (N-1) ] \,) \\
&= \frac{1}{\pi} \langle GS| c^{\dagger} \left[ \omega - (E_{GS} (N) - H_{tot}) - i \delta \right]^{-1} c|GS \rangle.
\end{split}
\label{eqn:XPS_rho}
\eeq
Here $c^{\dagger}$ is the creation operator of a core electron,  
$|GS \rangle$ and $| n (N-1) \rangle$ are, respectively, the ground state without a core hole, 
and eigenstates with a core hole \cite{CoreLevel, Hufner,PhysRevB.28.4315}. 
Once $H_{tot}$ is specified, the second line of Eq.~\eqref{eqn:XPS_rho} is used to 
compute the XPS spectra.
%Eq.~\eqref{eqn:XPS_rho} is used to compute XPS spectrum for a given model Hamiltonian $H_{tot}$  \cite{CoreLevel, Hufner,PhysRevB.28.4315}.
%From Eq.~\eqref{eqn:XPS_rho} are noted. 
Note that $\rho(\omega)$ is non-zero only when  $\omega = E_{GS} (N) -  E_{core} (N-1)$. 
What the XPS spectrum reflects is the core-hole energy spectrum weighted by the matrix element $|\langle n (N-1) | c |GS \rangle|^2$.
The XPS spectrum is also routinely plotted as a function of binding energy 
$E_B$, defined as $E_B \equiv h \nu - \phi - E_{kin} =  -\omega$ \cite{binding}. 
For the purpose of this work, the constant energy shift is not important and we  
focus only on the dependence of the spectrum on the ``relative binding energy'' or ``relative kinetic energy''.

Conventionally, one distinguishes between the initial-state and final-state effects in the XPS spectrum \cite{Hufner, CoreLevel}. 
For the initial-state effect [Fig.~\ref{fig:XPS_illu}(b)], the valence electrons are not affected by the created core hole. 
In this case the XPS peak position is determined by the core-level energy $\epsilon_c$ only. 
Within this scenario, any observed multi-peak structure in the measured XPS spectrum implies that targeted ions (where the photoelectrons are ejected from)
experience different environments within the same sample. 
For example at the Si/SiO$_2$ interface, the observed multiple peaks in the Si $2p$ spectrum, which corresponds to different 
Si oxidation states (from Si$^{0+}$ to Si$^{4+}$), are used to deduce and quantify the formation of SiO$_x$ at the interface  \cite{PhysRevB.38.6084}.
For the final-state effect [Fig.~\ref{fig:XPS_illu}(c)], the valence electrons do feel and respond to the potential caused by the creation of a core hole. 
In this case a spatially uniform system can also lead to  additional peak structure around $\epsilon_c$ in the XPS spectrum. 
A classic example is CeNi$_2$, which is a nominally $f^0$ material but displays three XPS peaks (from Ce $3d$ core level),
identified as $f^0$, $f^1$, $f^2$ \cite{PhysRevB.27.7330}. 
It was realized by Kotani and Toyozawa \cite{Kotani_1974, Kotani_99}, and by Gunnarsson and Sch\"onhammer \cite{PhysRevB.28.4315} that 
the multiple peaks in this material originate from the final-state effect, where the valence electrons response to the presence of a core hole, 
especially the core-hole-induced energy change of Ce $4f$ levels, plays an important role.
Simply put, for the initial-state effect, the ions of different nominal charges {\em preexist} in the sample;
for the final-state effect, the ions of different nominal charges are {\em created} after the applying photons produce core holes.
We believe the experimentally observed multi-peak structure in nominally $d^1$ and $n$-doped $d^0$ transition metal oxides should be understood 
as being due to the final-state effect. In the following we shall provide our experimental and theoretical analysis 
that leads to this conclusion. The key parameter determining the relative importance between
initial-state and final state effects will be discussed in Section IV.E.

\section{Experiments and key features}
In order to properly analyze the intrinsic XPS spectra of nominally $d^1$ transition metal oxides, 
we need to be able to grow single phase, crystalline layers of these materials and then measure their XPS 
spectra without exposing the samples to air, as these materials are not thermodynamically stable in the ambient and will slowly oxidize. 
The samples of  NbO$_2$, SrVO$_3$, and LaTiO$_3$, as well as SrTiO$_3$ with several dopants, 
are grown in a molecular beam epitaxy (MBE) chamber and 
then transferred {\em in situ} to a high resolution photoemission chamber. 
The two chambers are connected by an ultrahigh vacuum transfer line with a base pressure of $<1 \times 10^{-9}$ Torr, 
allowing for sample transfer between the growth and analysis chamber within 5 min. 
The photoemission chamber consists of a monochromated Al K$\alpha$ photon source ($h \nu$ = 1486.6 eV) and a VG Scienta R3000 analyzer. 
XPS spectra of the valence band, O $1s$, Nb $3d$, V $2p$, Ti $2p$, Sr $3d$, and La $3d$ are taken (as appropriate) 
at a pass energy of 100 eV with an analyzer slit setting of 0.4 mm, resulting in an overall instrumental resolution of 350 meV 
(primarily limited by the energy resolution of the x-ray source). The 
analyzer is calibrated such that the Fermi level of a clean silver foil is at 
a binding energy of 0.00 eV and the Ag $3d_{5/2}$ core level is at 368.28 eV. 

Undoped SrTiO$_3$ is nominally $d^0$, while the remaining three materials are nominally $d^1$: 
in the ionic limit, SrTiO$_3$ has no electron occupying the Ti $3d$ orbital;
NbO$_2$ has one electron occupying the Nb $4d$ orbital;
SrVO$_3$ and LaTiO$_3$ have one electron occupying the V $3d$ and Ti $3d$ orbital, respectively. 
NbO$_2$ films are grown on 111-oriented SrTiO$_3$ substrates as described in more detail elsewhere \cite{Posadas_NbO2.APL}. 
Both SrVO$_3$ and LaTiO$_3$ films are grown on 100-oriented SrTiO$_3$ substrates at a temperature of 600-800$^{\circ}$C 
using co-deposition of matched metal fluxes in the presence of between $3\times 10^{-9}$ to $2\times 10^{-8}$ 
Torr of molecular oxygen with a total growth rate of $\sim$0.4 nm/min. 
All films reported here are crystalline as-deposited, with pseudo-rutile structure for NbO$_2$ \cite{Posadas_NbO2.APL} 
and perovskite structure for SrVO$_3$ and LaTiO$_3$, as determined by reflection high energy electron diffraction (RHEED). 
We systematically vary the oxygen pressure during growth to 
determine the conditions that would result in the ideal O:Nb, O:Ti, and O:V ratios in the films. 
The transition metal to oxygen ratios are  determined by the integrated intensities of the relevant XPS core level spectra 
(O $1s$ for oxygen) and the appropriate atomic sensitivity factors, 
as well as verifying that the Sr:V and La:Ti ratios are very close to one. 
The atomic sensitivity factors used are empirical values as reported by Wagner et al. 
\cite{Wagner.ASF2, Wagner.ASF} and adjusted to give ideal oxygen to metal ratios for the compounds 
Nb$_2$O$_5$, V$_2$O$_5$, and undoped SrTiO$_3$. 
In the following, we present our experimental results for the transition metal core level spectra for
 single phase, nominally $d^1$ materials, and for $n$-doped SrTiO$_3$, as measured using {\em in situ} XPS. 
All materials are sufficiently conductive at room temperature such that there is negligible ($<0.1$ V) 
sample voltage during the measurement. For each material, we show a core level spectrum for an under-oxidized, optimally oxidized, 
and over-oxidized sample for comparison. The detailed results for each material are presented in the following sections.
In the Supplementary Material \cite{Supplementary}, we provide RHEED data for stoichiometric 
SrVO$_3$, LaTiO$_3$, and NbO$_2$ to further demonstrate our sample quality.

\subsection{NbO$_2$}
\begin{figure}[http]
\epsfig{file=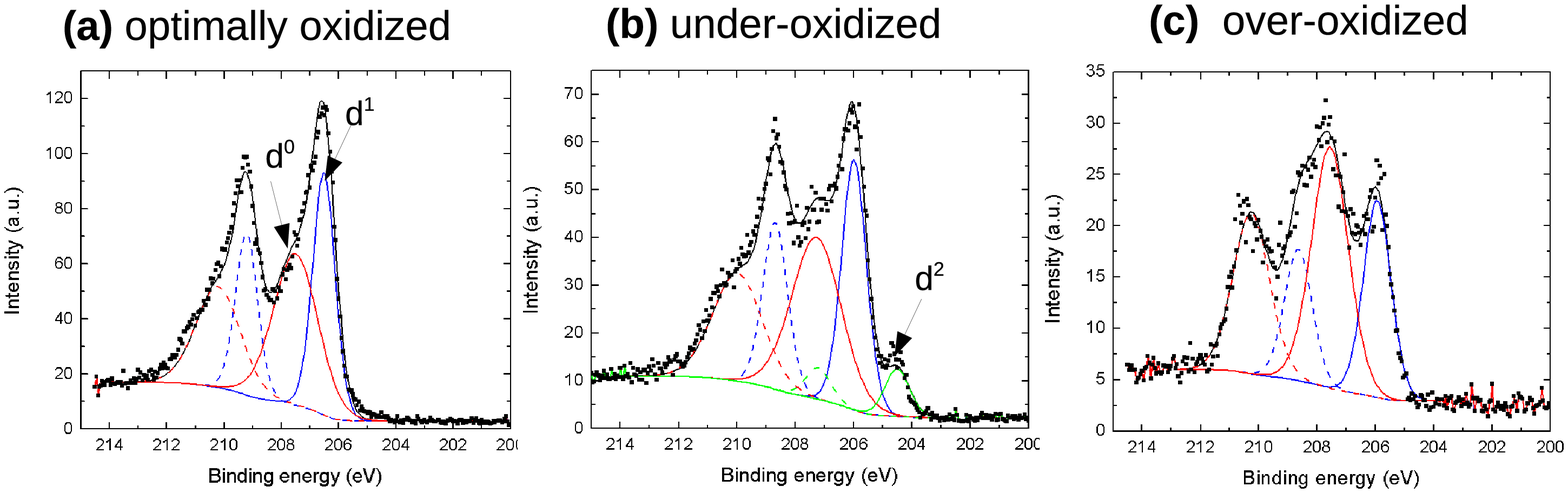, width = 0.96\textwidth}
\caption{ (Color Online) Nb $3d$ XPS spectrum of NbO$_2$:  (a) optimally oxidized;
(b) under-oxidized; (c) over-oxidized. The arrows indicate the $d^0$, $d^1$ and $d^2$ peaks.
The $d^0$/$d^2$ peak is most pronounced in over-oxidized/under-oxidized samples.
}
\label{fig:NbO2}
\end{figure}

The Nb $3d$ core level in Nb$_2$O$_5$ is located at a binding energy of 207.7 eV (Nb 3$d_{5/2}$) 
and has a spin-orbit pair at 2.7 eV higher binding energy (Nb $3d_{3/2}$). 
To model the Nb $3d$ multi-peak structure in NbO$_2$, 
we assume that the spin-orbit pairs are of the same width and that their separation is the same as in 
Nb$_2$O$_5$. Two or three pairs of peaks (pseudo-Voigt line shape) are used as needed to fit the data. 
For the optimally oxidized case (O/Nb = 2.0) shown in Fig.~\ref{fig:NbO2} (a),
we find two components. The first component has a binding energy of 206.5 eV with a width of 0.9 eV, 
while the second component has a binding energy of 207.5 eV and a larger width of 1.8 eV. 
If we assign the 207.5 eV feature to be the $d^0$ component, the $d^0$ component is 55\% of the integrated intensity 
while the $d^1$ component is 45\% of the integrated intensity.

If we electron dope the system by removing oxygen to form an under-oxidized NbO$_2$ phase [Fig. ~\ref{fig:NbO2} (b)] 
with O/Nb = 1.9, we find that both the $d^0$ component at 207.3 eV and the $d^1$ component at 206.0 eV decrease 
slightly in relative amount to 52\% and 40\% of the signal. A new component ($d^2$) emerges at a binding energy 
of 204.5 eV with a relative amount of 8\%. On the other hand, if we add excess oxygen to the system 
and form over-oxidized NbO$_2$ [Fig. ~\ref{fig:NbO2} (c)] with O/Nb = 2.1, the shape of the spectrum changes qualitatively. 
The $d^0$ component (at 207.6 eV) becomes sharper (width of 1.5 eV) and increases to 62\%, 
while the $d^1$ component at 205.9 eV (width of 1.1 eV) drops to 38\%.

\subsection{SrVO$_3$}
For SrVO$_3$, we look at the V $2p$ core level. For comparison, in pure V$_2$O$_5$, the V $2p_{3/2}$ peak is located 
at a binding energy of 517.9 eV, with the $2p_{1/2}$ spin-orbit pair located at 7.4 eV higher binding energy. 
The $2p_{1/2}$ peak is significantly broader than the $2p_{3/2}$ peak due to Coster-Kronig transitions. 
To model V $2p$ spectra, the widths of all $2p_{3/2}$ components are constrained to be the same and the widths of 
all $2p_{1/2}$ peaks are also constrained to be the same. There is no restriction on the relative 
widths of the $2p_{3/2}$ and $2p_{1/2}$ peaks within each component, however. The $2p_{3/2}$ to $2p_{1/2}$ separation of each component 
is also fixed to be the same as that of V$_2$O$_5$. Three sets of spin-orbit pairs of peaks are used to fit all the SrVO$_3$ data. 
Because the O $1s$ core level is near the V $2p$ levels, O $1s$ signals are also collected in the same measurement and included in the fitting. 

\begin{figure}[http]
\epsfig{file=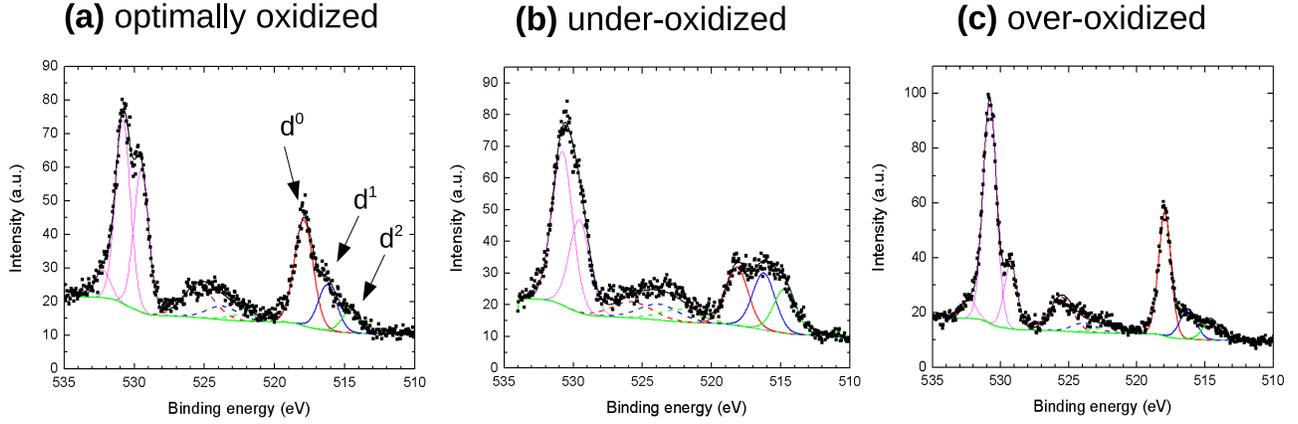, width = 0.96\textwidth}
\caption{ (Color Online) V $2p$ XPS spectrum of SrVO$_3$:  (a) optimally oxidized;
(b) under-oxidized; (c) over-oxidized. The arrows indicate the $d^0$, $d^1$ and $d^2$ peaks.
The $d^0$/$d^2$ peak is most pronounced in over-oxidized/under-oxidized samples.
We note that in SrVO$_3$, the optimally oxidized sample already displays a $d^2$ peak.
}
\label{fig:SrVO3}
\end{figure}

For the optimally oxidized case with O/V = 3.0 [Fig.~\ref{fig:SrVO3} (a)], 
the spectrum consists of three distinct components. 
The widths of the $2p_{3/2}$ peaks are 1.5 eV. The first peak has a binding energy of 517.9 eV ($d^0$) 
with a relative concentration of 60\%. The second peak ($d^1$) has a binding energy of 516.2 eV 
with a relative concentration of 27\%. The third peak ($d^2$) has a binding energy of 514.5 eV 
with a relative concentration of 13\%. Reducing the O/V ratio to 2.7 [Fig.~\ref{fig:SrVO3} (b)] results in a significant 
decrease in the $d^0$ component at 518.1 eV to 36\%. The $d^1$ component at 516.2 eV increases to 36\% 
while the $d^2$ component at 514.7 eV increases to 28\%. 
On the other hand, slightly over-oxidizing the SrVO$_3$ to have an O/V ratio of 3.1 [Fig.~\ref{fig:SrVO3} (c)] 
alters the relative amounts of the three components to 65\% for $d^0$, 22\% for $d^1$, and 13\% for $d^2$. 

\subsection{LaTiO$_3$}
\begin{figure}[http]
\epsfig{file=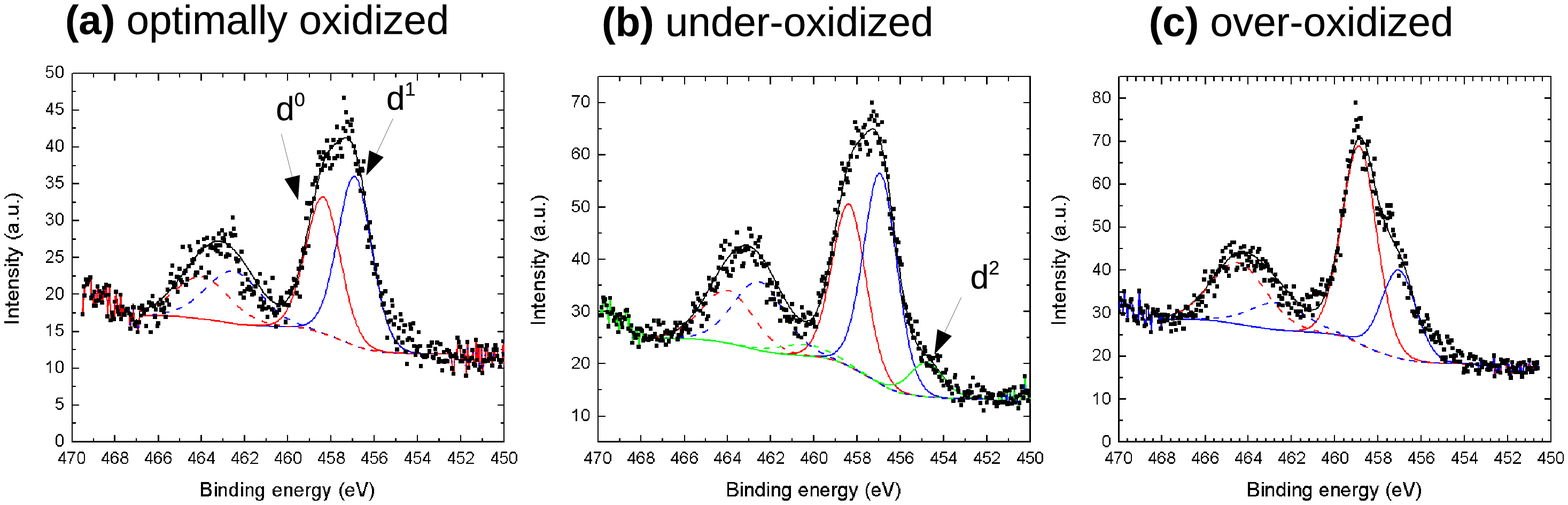, width = 0.96\textwidth}
\caption{ (Color Online) Ti $2p$ XPS spectrum of LaTiO$_3$:  (a) optimally oxidized;
(b) under-oxidized; (c) over-oxidized. The arrows indicate the $d^0$, $d^1$ and $d^2$ peaks.
The $d^0$/$d^2$ peak is most pronounced in over-oxidized/under-oxidized samples.
}
\label{fig:LaTiO3}
\end{figure}

For LaTiO$_3$, we use the Ti $2p$ core level. The Ti $2p_{1/2}$ level is significantly wider 
than the $2p_{3/2}$ level due to Coster-Kronig transitions. We model the Ti $2p$ spectra 
using the same kind of constraints on widths and spin-orbit separation as in the V $2p$ modeling. 
For comparison, the Ti $2p_{3/2}$ level of stoichiometric SrTiO$_3$ (Ti$^{4+}$) is located at 458.9 eV 
with a $2p_{3/2}$ to $2p_{1/2}$ separation of 5.6 eV. Two or three pairs of peaks are used to model the LaTiO$_3$ Ti $2p$ spectra as needed. 
For the optimally oxidized sample with O/Ti = 3.0 [Fig.~\ref{fig:LaTiO3} (a)], there are two components. 
The first one ($d^0$) is located at a binding energy of  458.4 eV with a relative concentration of 45\%. 
The second component ($d^1$) is located at a binding energy of 456.9 eV with a relative concentration of 55\%. 
The widths of both $2p_{3/2}$ peaks is 1.7 eV. 

When LaTiO$_3$ is under-oxidized to yield an O/Ti ratio of 2.8 [Fig.~\ref{fig:LaTiO3} (b)], 
we see the emergence of a third component ($d^2$) with a binding energy of 454.9 eV and a relative amount of 8\%. 
The other two components are both slightly reduced in amount to 40\% for $d^0$  and 52\% for $d^1$. 
For the slightly over-oxidized case, with O/Ti = 3.1 [Fig.~\ref{fig:LaTiO3} (c)], 
we see a significant increase in the $d^0$ component to 69\% 
with a slight shift in binding energy to 458.9 eV. The $d^1$ component (at binding energy 457.0 eV) correspondingly decreases to 31\%.

\subsection{$n$-doped SrTiO$_3$}

\begin{figure}[http]
\epsfig{file=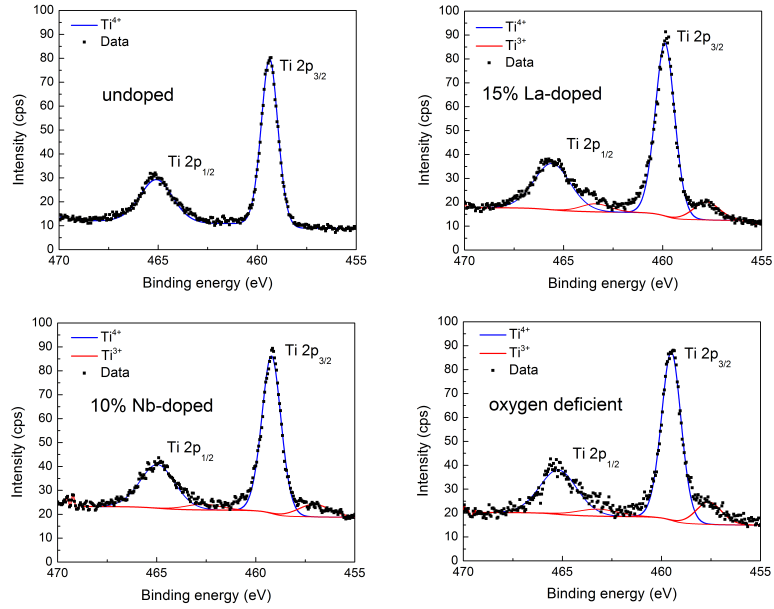, width = 0.7\textwidth}
\caption{ (Color Online) Experimental Ti $2p$ XPS spectra for stoichiometric, 15\% La-doped, 10\% Nb-doped, 
and oxygen-deficient SrTiO$_3$.
A shoulder, labeled as Ti$^{3+}$, at about 1.5 eV below the main Ti$^{4+}$ peak appears for all $n$-doped samples.
The dots are experimental data and the solid curves are from Gaussian fitting.
The experimental details are given in Ref.~\cite{/content/aip/journal/jap/116/4/10.1063/1.4891225}.
}
\label{fig:XPS_STO}
\end{figure}

In stoichiometric SrTiO$_3$, the $2p_{3/2}$ peak shows a single feature about 1 eV wide with no shoulder \cite{note1}.
The $2p_{1/2}$ peak is significantly broader than the $2p_{3/2}$ peak due to Coster-Kronig transitions \cite{Coster_Kronig}. 
Fig.~\ref{fig:XPS_STO} shows the Ti $2p$ XPS spectra for 15\% La doped SrTiO$_3$ (Sr$_{1-x}$La$_{x}$TiO$_3$) 
\cite{PhysRevLett.70.2126,/content/aip/journal/apl/100/26/10.1063/1.4731642, /content/aip/journal/jap/116/4/10.1063/1.4891225}, 
10\% Nb doped SrTiO$_3$ (SrTi$_{1-x}$Nb$_{x}$O$_3$) \cite{PhysRevB.61.12860}, 
and  oxygen-deficient SrTiO$_3$ (SrTiO$_{3-x}$) \cite{NatMetal_STO_2D, Hatch_13, Rice_2014}. 
In all these $n$-doped SrTiO$_3$, a small shoulder located about 1.5 eV lower than the Ti$^{4+}$ peak emerges, 
and is typically interpreted as a Ti$^{3+}$ ($d^1$) peak.
Two important features should be pointed out. First, the position and strength of Ti$^{3+}$ peak are not sensitive 
to photoelectron emission angle (not shown), indicating that this signal is not a surface effect.
Second, the position of Ti$^{3+}$ peak is dopant-independent, indicating that 
this peak is very likely to be intrinsic to doped SrTiO$_3$.
We will show in the next section that these two observations are consistent with 
the final-state interpretation. 

\subsection{Common features of $d^1$ transition metal oxide spectra}
We summarize this section by pointing out the key common features of the XPS spectra of these $d^1$ transition metal oxides:
the transition metal core level spectra of these materials all display at least two, and sometimes three distinct components 
(where a component refers to a pair of peaks related by spin-orbit coupling); 
a single component is never observed even in the optimally oxidized single phase films.
These XPS peaks can be assigned as $d^0$ (Nb$^{5+}$, V$^{5+}$, Ti$^{4+}$), 
$d^1$ (Nb$^{4+}$, V$^{4+}$, Ti$^{3+}$), and $d^2$ (Nb$^{3+}$, V$^{3+}$, Ti$^{2+}$) oxidation states.
As a general trend, electron doping (via oxygen vacancies) increases the intensity of 
the $d^2$ peak at the expense of the $d^0$ and $d^1$ peaks, whereas 
hole doping (via oxygen excess) increases that of the $d^0$ peak and decreases the intensity of the $d^2$ peak if present.
Based on the initial-state effect, one might naively infer from the XPS results 
that the optimally oxidized samples contain significant amounts of regions of different 
oxidation states (such as Nb$_2$O$_5$ which is nominally d$^0$).
However, this interpretation is not consistent with RHEED from the samples, 
which should clearly show the presence of incommensurate monoclinic/amorphous Nb$_2$O$_5$ 
or pyrochlore La$_2$Ti$_2$O$_7$/Sr$_2$V$_2$O$_7$ phases, if they are present in such large amounts. 
Quantitatively, if we assume the peak intensity of a particular component is proportional to the abundance of 
that particular oxidation state, this implies that roughly one half of the sample on average is in the highest oxidation state. 
For example, from the XPS of SrVO$_3$ [Fig.~\ref{fig:SrVO3}], one expects 60\% of the sample to consist of pyrochlore Sr$_2$V$_2$O$_7$ 
which should be, but is not, reflected in the diffraction data, which still shows a single phase, epitaxial 100-oriented pervoskite film. 
It should also be noted that the oxygen to transition metal ratio has been carefully controlled during growth (as described above),
spanning the range from under-oxidized to over-oxidized.
Furthermore, we also note that growing at very low oxygen pressures 
that result in an oxygen to metal ratio significantly less than the ideal value still results in the presence of 
a peak that is associated with the $d^0$ oxidation state. 
The presence of a strong $d^0$ peak in stoichiometric SrVO$_3$ has been interpreted by Takizawa et al.
\cite{PhysRevB.80.235104, Takizawa_thesis} as being due to excess oxygen (forming V$^{5+}$) decorating the surface of SrVO$_3$
resulting in a $\sqrt{2}\times \sqrt{2}$ reconstruction pattern.  As shown in the Supplementary Materials \cite{Supplementary}, 
 we also observe the surface reconstruction in RHEED. By comparing the XPS spectra before and after the
 Ar sputtering (which removes the surface atoms), we conclude that both 
the surface reconstruction (initial-state effect) and final-state effect contribute to the multi-peak structure in the case of SrVO$_3$.
In a vacuum-cleaved single crystal of SrVO$_3$, only a weak $d^0$ feature is observable \cite{Eguchi2007421}. 
The seemingly conflicting results from the XPS data and the single phase nature of the optimally oxidized films can be naturally reconciled 
if the occurrence of the multi-peak structure in the XPS spectra is intrinsic to these $d^1$ materials (i.e.
the spatially uniform $d^1$ system by itself displays multiple peaks in XPS). 
In the next section we argue that it is indeed the case once the final-state effect is considered, and 
provide a simple model to illustrate this point.

\section{Model and theoretical analysis}

\subsection{Model and parameters}
To explain the observed multi-peak structure in the XPS spectra, we propose a cluster-bath model 
which resembles that proposed in Ref.~\cite{PhysRevB.76.035101, PhysRevB.78.075103}.
It contains three parts [see Fig.~\ref{fig:ClusterXPS_bath_cubic}(a)]:
\beq 
H_{tot} = H_{cluster} + H_{bath} + H_{cl-bath}.
\label{eqn:H_tot}
\eeq 
$H_{cluster}$ describes a TM-O$_6$ (TM can be Ti, V, or Nb) cluster that
includes at least five TM $d$ and three O $2p$ orbitals (for each of six oxygen atoms).
We point out that our model does not qualitatively distinguish between the $3d$ and $4d$ orbitals, or $3d$ 
orbitals of different chemical elements: they only correspond to different parameters in the model.
When taking the cubic symmetry into account, only ten of  twenty-three total orbitals couple to one another \cite{12_orbital}.
Using $\Gamma$ to label the orbital symmetry (three $t_{2g}$ $xy$, $yz$, $zx$ and two $e_g$ $3z^2-r^2$, $x^2-y^2$ orbitals),
the cluster Hamiltonian \cite{Cluster_note} is
\beq 
\begin{split}
 H_{cluster} &= \sum_{\Gamma,\sigma} \left\{ \epsilon_p (\Gamma) n_{p, \Gamma,\sigma}
 + \epsilon_d (\Gamma) n_{d, \Gamma,\sigma}
 + V(\Gamma) [d^{\dagger}_{\Gamma,\sigma} p_{\Gamma,\sigma} + H.c.] \right\} \\
 &+ \frac{U}{2} \sum_{(\Gamma, \sigma) \neq (\Gamma', \sigma')} n_{d, \Gamma,\sigma} n_{d, \Gamma',\sigma'}
 - U_{dc} (1-n_{core}) \sum_{\Gamma, \sigma} n_{d, \Gamma, \sigma} + \epsilon_c n_{core}.
\label{eqn:dp_cluster}
\end{split}
\eeq
Here $n_{d, \Gamma,\sigma} = d^{\dagger}_{\Gamma,\sigma} d_{\Gamma,\sigma}$, 
$n_{p, \Gamma,\sigma} = p^{\dagger}_{\Gamma,\sigma} p_{\Gamma,\sigma} $ are respectively the TM $d$ and O $2p$ number operators 
for the orbital labeled by $(\Gamma, \sigma)$ ($\sigma$ labels the spin). 
$\epsilon_d (\Gamma) $ and $\epsilon_p (\Gamma)$ are energies of  
TM $d$ and O $2p$ orbitals, and $V(\Gamma)$ describes their hybridizations.
$U$ is the energy cost when the $3d$ occupation of the TM atom is more than one.
$n_{core}$ is the number operator of the core level and $\epsilon_c$  the core-level energy 
(approximately $-459.0$ eV for Ti $2p$, $-518.0$ eV for V $2p$,  $-208.0$ eV for Nb $3d$).
The term with $U_{dc}$ approximates how valence electrons 
respond to the core hole: in the presence of a core hole (i.e. $\langle n_{core} \rangle = 0$), all TM $d$ levels are shifted down
by $U_{dc}$ to screen the core hole. 
By fitting to published experimental data from  XPS of SrTiO$_3$, ellipsometry, and angle-resolved photoemission spectroscopy (ARPES) 
\cite{Kotani_93, Zollner, J.Appl.Phys.90.6156, Hatch_13, PhysRevB.79.113103},
we take $\epsilon_d (e_g) = 2.0 $ eV,  $\epsilon_d (t_{2g}) = 0 $ eV, $\epsilon_p (\Gamma) = -3.0$ eV, 
$V (e_g) = 2.5 $eV,  $V (t_{2g}) = -1.3$ eV, $U=6.0$ eV, and $U_{dc} = 8.0$ eV \cite{Fitting}.
This problem can be solved exactly by the technique introduced by 
Gunnarsson and Sch\"onhammer \cite{PhysRevB.28.4315, key_trick}, and the details 
are provided in Appendix B.

For the metallic phase, the occupation of each local orbital fluctuates.
To capture this effect, we further introduce a set of bath orbitals, which 
simulate the role of TMO conduction bands, coupling to each $d$ orbital.
For Hamiltonians involving the bath:
\beq 
\begin{split}
H_{bath} + H_{cl-bath} &= \sum_{\Gamma,\sigma}
\left[  \int \epsilon \,b^{\dagger}_{\epsilon \Gamma \sigma} b_{\epsilon \Gamma \sigma} d\epsilon
+ \int [V(\epsilon, \Gamma) d^{\dagger}_{\Gamma \sigma} b_{\epsilon \Gamma \sigma} +H.c.]d\epsilon,
\right] \\ 
\pi |V(\epsilon, \Gamma)|^2 &= \frac{2V^2}{B^2} \sqrt{B^2 - (\epsilon-\epsilon_0)^2} d\epsilon.
\end{split}
\label{eqn:bath_coupling}
\eeq
Here $ b_{\epsilon \Gamma \sigma}$ denotes the bath orbitals 
of energy $\epsilon$, orbital symmetry $\Gamma$ and spin $\sigma$.
Inclusion of the bath introduces charge fluctuation in the cluster (via
exchange of particles with the bath) that is used to model 
the fluctuation in the occupation of local orbital  in the metallic phase \cite{PhysRevB.28.4315}
(see Appendix A for a simple explanation). 
%in a metallic state, the occupation of a local orbital fluctuates \cite{PhysRevB.28.4315}. 
We use $\epsilon_0 = 2.0$ eV, $B = 2.0$ eV (so the bath levels range from 0 to 4.0 eV, roughly
the SrTiO$_3$ conduction bandwidth)
to approximate the SrTiO$_3$ conduction bands, and take $V=0.3$ eV which 
is approximately the effective hopping between two adjacent Ti $3d$ orbitals \cite{Zollner,PhysRevB.87.161102}. 
It turns out that the exact value of $V$ plays a relatively  minor role in the  XPS spectrum (see Appendix B).  
%for the cluster part; 
%$\epsilon_0 = 2.0$ eV, $B = 2.0$ eV and $V=0.3$ eV for terms involving the bath orbitals.
In the calculation, we introduce the chemical potential $\mu$ to specify the number of total electrons (filling) in the 
whole system (bath and cluster): all bath levels  below $\mu$ are filled.
Qualitatively larger $\mu$ corresponds to larger average $d$ occupation in the bulk material.
To extract the essential feature of these $d^1$ materials, 
we only vary $\mu$ but keep all other parameters fixed. In other words,  the valence levels of
Ti $3d$, V $3d$, and Nb $4d$ are not distinguished in our simulation. 
The XPS spectrum is calculated using Eq.~\eqref{eqn:XPS_rho}, and the details are given in the Appendix B.
%Essentially we first solve the ground state $|GS \rangle$, and then follow the Lanczos procedure 
%\cite{Grosso, RevModPhys.66.763} to compute the spectrum.

\subsection{Results from an isolated cluster}

\begin{figure}[ht]
\epsfig{file=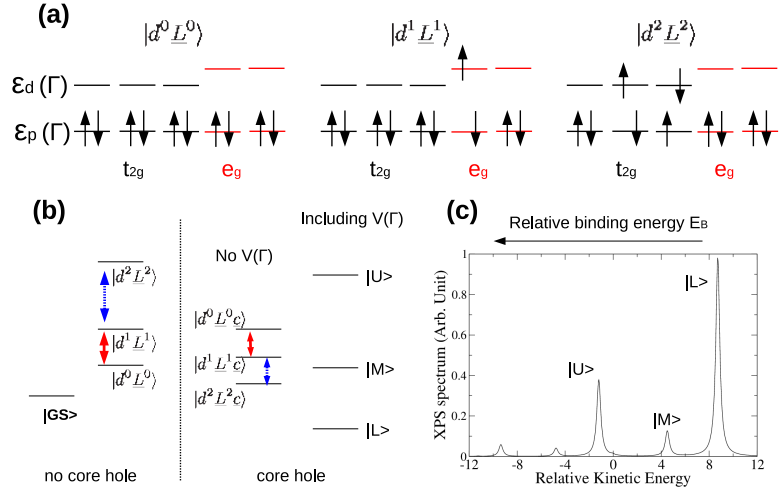, width = 0.6\textwidth}
\caption{ (Color Online) (a) The orbitals kept in the cluster.
For the state $|d^{0} \underline{L}^0\rangle$, all $2p$ levels are filled.
For the state $|d^{1} \underline{L}^1\rangle$, one of the $3d$ levels is filled, resulting in a $2p$ hole of the 
same orbital symmetry. For the state $|d^{2} \underline{L}^2\rangle$, two of the $3d$ levels are filled.
(b) A simple energy spectrum, keeping only the $|d^{0} \underline{L}^0\rangle$,
$|d^{1} \underline{L}^1\rangle$, $|d^{2} \underline{L}^2\rangle$ states. The double-arrow indicates 
the non-zero coupling between these states. The ground state $|GS \rangle$ is a linear combination of these three states. 
In the presence of a core hole (we add $\underline{c}$ to indicate its presence),
the relative energy levels change due to the screening, but their hybridizations stay the same.  
The resulting eigenstates, labeled as $|L \rangle$, $|M \rangle$ and $|U \rangle$, correspond 
to the peak-structure in the XPS spectrum.
(c) The computed XPS spectrum in arbitrary units, computed with a broadening of 0.2 eV. 
Three main peaks can be understood as the excitations 
in the presence of a core hole. The peak $|L\rangle$ is conventionally assigned as the $d^{0}$ peak
(in the sense of $|d^{0} \underline{L}^0\rangle$),
although it contains significant contributions from $|d^{1} \underline{L}^1\rangle$ and $|d^{2} \underline{L}^2\rangle$.
}
\label{fig:d-pcluster}
\end{figure}

Before discussing our results using the total Hamiltonian Eq.~\eqref{eqn:H_tot}, 
we first present the results from the isolated cluster (zero impurity-bath coupling).
In particular we shall identify the origin of each peak.
Fig.~\ref{fig:d-pcluster}(c) shows the XPS spectrum of an isolated cluster -- ten electrons 
are filled to mimic the nominally $d^0$ system. 
There are three pronounced peaks, labeled as $|L \rangle$, $|M \rangle$, and $|U \rangle$ 
referring to their relative lower, middle, and upper binding energies.
These features can be understood by considering the following three states 
$|d^{0} \underline{L}^0\rangle$, $|d^{1} \underline{L}^1\rangle$, and $|d^{2} \underline{L}^2\rangle$ \cite{Kotani_93, CoreLevel}.
Here $|d^{0} \underline{L}^0\rangle$ represents the ``reference'' state where all O $2p$ orbitals are filled, and
$|d^{i} \underline{L}^i\rangle$ represents the state of $i$ particle-hole (p-h) pairs with respect to 
$|d^{0} \underline{L}^0\rangle$ 
[see Fig.~\ref{fig:d-pcluster}(a) for illustration].
Without the core hole, the ground state $|GS \rangle$ is a linear combination of these three states.
States with larger number of p-h pairs are significantly less important due to the on-site energy $U$.
In the presence of a core hole (we use $|d^{i} \underline{L}^i \underline{c}\rangle$ 
to denote states in the presence of a core hole),
the relative energies of these three states change, and the resulting core-hole eigenstates (including 
the d-p hybridization) are labeled as $|L \rangle$, $|M \rangle$, $|U \rangle$. These three lowest eigenstates account for  
the three pronounced peaks in the computed spectrum. From Eq.~\eqref{eqn:XPS_rho}, the peak strength is given by 
$|\langle X| c |GS \rangle|^2$ for $X=L,M,U$.
We emphasize that, due to the strong d-p hybridization,
all core-hole eigenstates $|L \rangle$, $|M \rangle$, $|U \rangle$ 
have significant $|d^{i} \underline{L}^i \underline{c}\rangle$ ($i=0,1,2$) components.
Comparing with the experimentally observed XPS SrTiO$_3$ spectrum [Fig.~\ref{fig:XPS_STO}], we note that: (i)
the strongest peak $|L\rangle$ is conventionally assigned as the Ti$^{4+}$ ($d^0$) $2p_{3/2}$ peak;
(ii) the weak peak $|M \rangle$ is buried under the $2p_{1/2}$ peak  caused by the 
spin-orbit coupling of the core electron, and is not observed;
(iii) the calculated $|U\rangle$ peak  corresponds to the charge transfer satellite feature at a binding energy 
of approximately 471.0 eV \cite{Kotani_93,Comparison_TiO2_STO} and
appears to be much sharper than that in the experiment, 
because we neglect the coupling between valence electrons and the core spin
that provides additional decay channels for states of higher binding energies \cite{Kotani_93,core_spin_broadening}.
In the following discussion we only focus on the strongest and lowest  peak,
which is the one used to determine the different oxidation states. 

%(ii) the weak peak $|M \rangle$ is buried under the $2p_{1/2}$ peak [Fig.xx] caused by the 
%spin-orbit coupling of the core electron, and is not observed;
%(iii) the calculated $|U\rangle$ peak  corresponds to the charge transfer satellite feature at a binding energy 
%of approximately 471.0 eV \cite{Kotani_93, Comparison_TiO2_STO} and
%appears to be much sharper than that in the experiment, 
%because we neglect the coupling between valence electrons and core spin,
%which provides additional decay channels to states of higher binding energies \cite{Kotani_93,core_spin_broadening}.

\subsection{Results including bath}

Inclusion of the bath introduces charge fluctuation in the cluster, as the cluster can now 
exchange particles with the bath orbitals (see Appendix A). More specifically, instead of 
the fixed number of electrons in the cluster, the total ground state wave function has a general form 
\beq 
|GS \rangle = \sum_{i=0} \alpha_i |n_{cl} + i \rangle_{cl} \otimes |n_{b} - i \rangle_{bath} \otimes |1 \rangle_{core}.
\label{eqn:particle_fluctuation}
\eeq
Here $|n_{cl} + i \rangle_{cl} \otimes |n_{b} - i \rangle_{bath} \otimes |1 \rangle_{core} $ represents a state 
which has $n_{cl} + i$ particles in the cluster, $n_{b} - i $ particles in the bath and a filled core level.
Using the same notation as Eq.~\eqref{eqn:particle_fluctuation}, the isolated cluster calculation presented 
in the previous subsection only has $|n_{cl}=10 \rangle_{cl} \otimes  |n_{b}  \rangle_{bath} \otimes |1 \rangle_{core}$, with
$|n_{cl}=10 \rangle_{cl}$ including all possible  $|d^{i} \underline{L}^i\rangle$ ($i$=0 to 10 in principle) 
components. When exchanging particles with the bath, the states such as $|n_{cl}=11 \rangle_{cl}$ ($|d^{i+1} \underline{L}^i\rangle$, 
$i =0$ to 9), $|n_{cl}=12 \rangle_{cl}$ ($|d^{i+2} \underline{L}^i\rangle$, $i =0$ to 8) 
also contribute to the $|GS \rangle$. Similarly, in the presence of a core hole,
the $n$th eigenstate with energy $E_{core,n} (N-1)$ has the general form 
\beq 
|n (N-1) \rangle = \sum_{i=0} \beta^{(n)}_i |n_{cl} + i \rangle_{cl} \otimes |n_{b} - i \rangle_{bath} \otimes |0 \rangle_{core}.
\label{eqn:particle_fluctuation_minus}
\eeq
Applying Eq.~\eqref{eqn:particle_fluctuation} and Eq.~\eqref{eqn:particle_fluctuation_minus} to Eq.~\eqref{eqn:XPS_rho}, 
the XPS spectrum displays peaks at $\omega_n = E_{GS} (N) -  E_{core,n} (N-1)$ with weight $|\sum_i \beta^{(n)}_i  \alpha_i |^2$. 
From this general analysis, we see that including the charge fluctuation  naturally leads to 
multiple XPS peaks, which correspond to different particle number in the cluster.

\begin{figure}[ht]
\epsfig{file=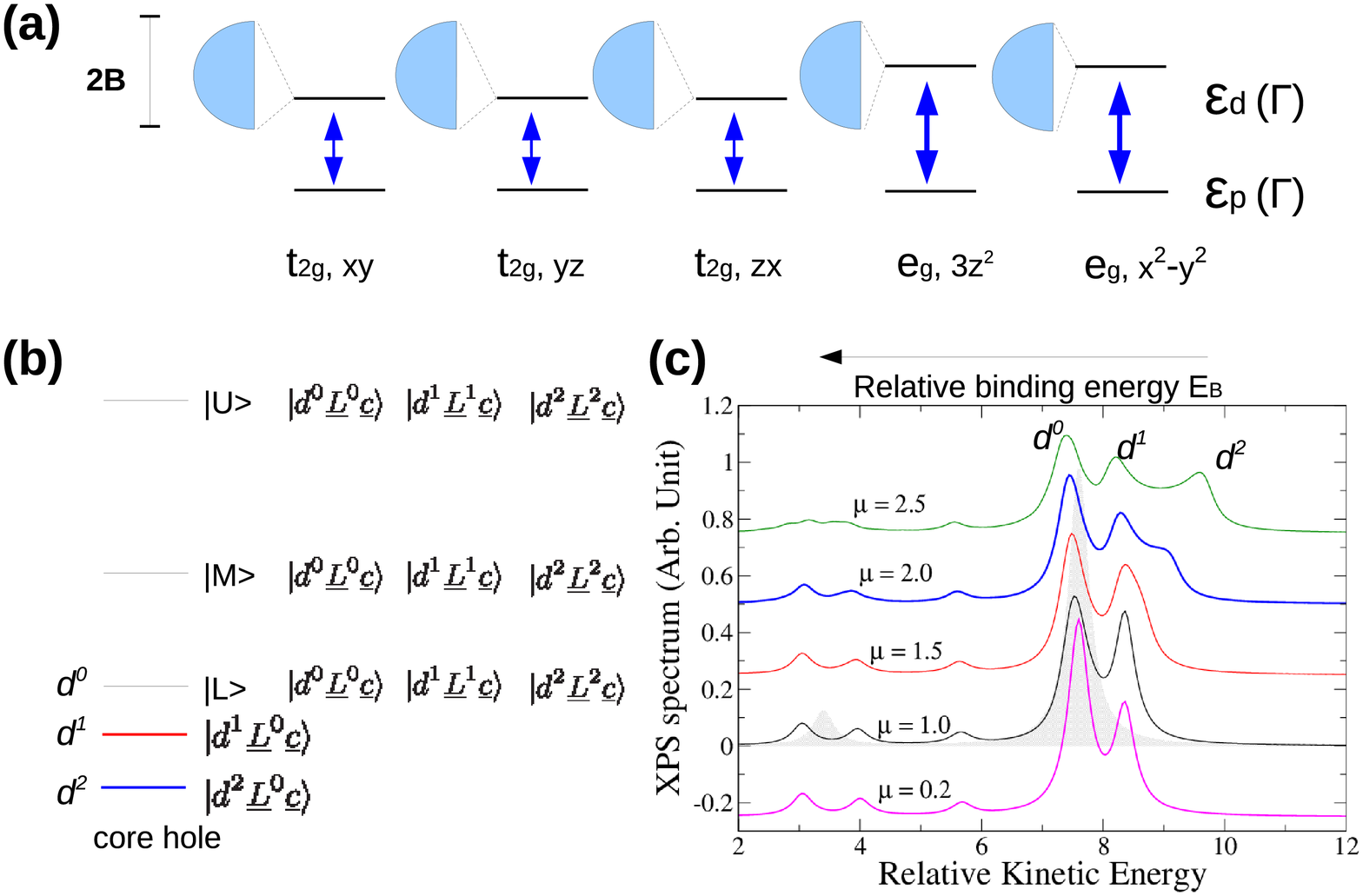, width = 0.6\textwidth}
\caption{ (Color Online) 
(a) Schematic illustration of the cluster impurity model.
Each $d$-orbital couples to a set of uncorrelated bath orbitals.
(b) Cluster energy spectrum with a core hole. 
Due to the bath coupling, the number of electrons within the cluster is not a constant.
In particular, the state of $|d^1 \underline{L}^0 \underline{c} \rangle$, $|d^2 \underline{L}^0 \underline{c} \rangle$ 
accounts for the $d^1$, $d^2$ peaks at lower binding energy.
(c) The computed XPS spectra (arbitrary units) for $\mu=0.2, 1.0, 1.5, 2.0, 2.5$ eV.
The spectrum of $\mu=0.0$ is given as the shaded region.
Increasing the  chemical potential increases the intensity of $d^1$ and $d^2$ peaks and 
reduces the $d^0$ peak. For these chemical potentials both $d^0$ and $d^1$ peaks always exist.
At $\mu=2.0$ eV, the $d^2$ peak begins to emerge as a shoulder at the lower binding energy. 
A Lorentz broadening of 0.2 eV is used. 
}
\label{fig:ClusterXPS_bath_cubic}
\end{figure}

In Fig.~\ref{fig:ClusterXPS_bath_cubic}(c) we show the calculated XPS spectra for $\mu=0.2, 1.0, 1.5, 2.0$, and 2.5 eV.
Starting from the highest chemical potential, the $\mu=2.5$ eV XPS spectrum shows three distinct peaks.
By analyzing the wave functions, they correspond to $|n_{cl}=10 \rangle_{cl}$, 
$|n_{cl}=11 \rangle_{cl}$ and $|n_{cl}=12 \rangle_{cl}$ in Eq.~\eqref{eqn:particle_fluctuation_minus}, 
and are therefore labeled as $d^0$, $d^1$, $d^2$ respectively.
Using the notation within the isolated cluster, the $d^0$, $d^1$ and $d^2$ peaks come from 
states of $|L\rangle$ ($\in |n_{cl}=10 \rangle_{cl}$), $|d^1 \underline{L}^0 \underline{c} \rangle $ ($\in |n_{cl}=11 \rangle_{cl}$),  
$|d^2 \underline{L}^0 \underline{c} \rangle$  ($\in |n_{cl}=12 \rangle_{cl}$) respectively,
as shown in Fig.~\ref{fig:ClusterXPS_bath_cubic}(b).
Decreasing $\mu$ reduces the intensities of $d^2$ and $d^1$ peak but increases that of $d^0$.
This is because lowering the chemical potential decreases the probability of adding 
electrons to the cluster from the bath, resulting in a smaller $|n_{cl}=11 \rangle_{cl}$ and $|n_{cl}=12 \rangle_{cl}$
components in $|GS \rangle$ and consequently weaker $d^1$ and $d^2$ peak intensities.
However, we stress that once the cluster and bath can exchange particles, 
a single $d^1$ XPS peak is never obtained in our calculation; the $d^0$ peak is always present.

\subsection{Discussion}

\subsubsection{Comments on experiments}
We now discuss several experiments based on the calculation.
The main conclusion from our model calculation is that, once 
charge fluctuation is taken into account, the nominally $d^1$ or $n$-doped $d^0$ transition metal oxides are expected to display multiple peaks 
in their XPS spectra, even in the absence of other oxidation states.
In other words, our theory implies that a multi-peak structure in XPS is 
{\em general} for these materials if charge fluctuation cannot be neglected.
%, and we shall comment on this condition will be discussed shortly.

We first discuss three observations in $n$-doped SrTiO$_3$ samples based on the 
general consequences of the final-state interpretation. 
First, the Ti$^{3+}$ peak position is dopant independent and 
is an intrinsic property of the Ti atom, or more precisely the TiO$_6$ cluster. 
Indeed, in lightly $n$-doped SrTiO$_3$, the Ti$^{3+}$ peaks all appear in the same position 
relative to the Ti$^{4+}$ peak 
\cite{PhysRevB.83.035410, /content/aip/journal/jap/116/4/10.1063/1.4891225,/content/aip/journal/apl/100/26/10.1063/1.4731642}
(Fig.~\ref{fig:XPS_STO}). Special attention is paid to the Nb-doped SrTiO$_3$ (or Nb-doped TiO$_2$ \cite{PhysRevB.61.13445}), 
where even in the ionic limit, there can only be Nb$^{4+}$ ions (i.e. Nb keeps one $4d$ electron), 
but not Ti$^{3+}$ ions.
Within our interpretation, the Nb gives its $4d$ electron to the conduction band,
resulting in  a metallic state and  nominally Nb$^{(5-x)+}$ and Ti$^{(4-x)+}$ ions 
(instead of Nb$^{4+}$ and Ti$^{4+}$), with Ti$^{(4-x)+}$ ions providing the XPS Ti$^{3+}$ signal.
Second, the Ti$^{3+}$/(Ti$^{4+}$+Ti$^{3+}$) ratio is routinely used to estimate the dopant concentration, 
and gives very reasonable values, which are consistent with other experiments
such as Hall measurements and Rutherford backscattering for low to moderate doping
\cite{PhysRevB.83.035410, /content/aip/journal/jap/116/4/10.1063/1.4891225,/content/aip/journal/apl/100/26/10.1063/1.4731642}.
According to our theory, this is possible because the Ti$^{4+}$ is the highest 
oxidation state and contains only one main peak. The Ti$^{3+}$ signal 
therefore appears as an extra, distinct side peak when reducing the average Ti oxidation state via doping. 
For a nominally $d^1$ system (that will be discussed shortly), multi-oxidation peaks exist intrinsically in the first place,
 and doping does not introduce a new peak. 
Also, we expect that using the Ti$^{3+}$/(Ti$^{4+}$+Ti$^{3+}$) ratio always slightly underestimates 
the dopant concentration as the nominally pure  Ti$^{3+}$ material already has significant Ti$^{4+}$ signal.
This is consistent with the results in Ref.~\cite{/content/aip/journal/jap/116/4/10.1063/1.4891225}.
%%%%
Finally, one cannot really distinguish the initial-state and final-state effect based solely on the XPS spectrum.
Both spatially localized Ti$^{3+}$ ions or a uniformly distributed Ti$^{(4-x)+}$ can account for the XPS Ti$^{3+}$ peaks.
The key difference between these two scenarios is that the former implies the presence of
an in-gap state, whereas the latter does not. 
To differentiate between them, one should probe the valence states to 
see if there is an in-gap signal. In oxygen-deficient SrTiO$_3$, an in-gap 
signal is observed in ARPES \cite{OV_Arpes_2002, NatMetal_STO_2D, Hatch_13}.
In this case the XPS Ti$^{3+}$ peak can be due to the presence of  localized Ti$^{3+}$ ions.
We note that in the literature, an oxygen vacancy is suggested to be 
a single donor \cite{Hou_10, PhysRevLett.111.217601}, which would
result in nominally localized Ti$^{3.5+}$ ions (we favor this view). Within the final-state effect, 
Ti$^{3.5+}$ ions also lead to a separate XPS Ti$^{3+}$ peak.
 It is worth noting that in the LaAlO$_3$/SrTiO$_3$ interface, the oxygen vacancies are responsible for 
the majority of charge carrier \cite{PhysRevB.75.121404,PhysRevLett.102.176805}.
However, the x-ray absorption spectrum does not indicate the existence of Ti$^{3+}$ ions
\cite{PhysRevLett.102.166804,PhysRevLett.111.087204}.

For the nominally $d^1$ TMO, all the optimally oxidized $d^1$ samples we have grown (as well as
 vacuum-cleaved single crystal  Ti$_2$O$_3$ \cite{Comparison_TiO2_STO}), 
demonstrate the XPS spectra showing multiple components. By viewing the multi-component structure as being caused by the final-state effect,
the existence of these multiple components does not require the presence of
different oxidation states in the sample. 
Even though XPS data show multiple components, the systematic way in which the oxygen 
content is controlled in the growth experiments, in combination with the single phase 
RHEED patterns observed, precludes the existence of different oxidation environments in the optimally oxidized samples.
The final-state interpretation reconciles the seeming conflict between XPS data, 
the single phase pattern in RHEED measurements, as well as the careful, systematic way in which 
the oxygen content is controlled in these growth experiments, which precludes the existence of different oxidation environments.
%Our calculation is also consistent with the XPS spectra of Ti$_{1-x}$Nb$_x$O$_2$ measured in Ref.~\cite{PhysRevB.61.13445} [add other references for SrVO3 spectra].
Moreover, our calculation shows the same doping dependence of the relative peak intensities:
increasing the electron doping decreases the $d^0$ peak intensity and causes an increase in the intensity of the $d^2$ peak.
This qualitative agreement between theory and experiment leads us to believe that 
the multi-peak structure in the single phase $d^1$ transition metal oxides actually originates from the final-state effect and
is intrinsic.
Certainly, as mentioned previously, one cannot rule out the initial-state effect, and ions of higher oxidation states
($V^{5+}$ for example) may exist at or near the surface of the sample. 
As observed in some vanadates  \cite{Eguchi2007421,PhysRevB.80.235104, Takizawa_thesis}, these ions  
also result in $d^0$ signal. However, we notice that even if these ions do exist, the 
$d^0$ signals appear to be too strong ($d^0$ and $d^1$ peaks are of comparable strength) to be interpreted as being solely from them.
In fact, we believe in SrVO$_3$,  the surface reconstruction (initial-state effect) and 
final-state effect {\it both} contribute to the observed $d^0$ peak (see the Supplementary Materials \cite{Supplementary}).

\subsubsection{Limitations of the theory}
There are two uncertainties in our model which make a more quantitative analysis difficult.
First it is not easy to map the chemical potential $\mu$ to the average $d$ occupancy in the bulk material.
Second, the energy distribution of bath orbitals and the cluster-bath coupling are also hard to determine.
However, the multi-peak structure is insensitive to these uncertainties (see Appendix B).
Namely, as long as there are particle exchanges between the cluster and the bath, there are multiple peaks 
in the XPS spectrum. For this reason we believe the conclusions drawn from our model are qualitatively correct.

\subsubsection{Charge fluctuation}
We now discuss the origin and the importance of charge fluctuation. 
The charge fluctuation cannot be neglected in the metallic state, where particle exchange with
the Fermi sea causes fluctuation in the occupation of local orbitals \cite{PhysRevB.28.4315}.
Accordingly, charge fluctuations in doped or metallic samples should not be neglected,
and multiple XPS peaks in these samples are expected (and indeed observed) \cite{PhysRevB.61.13445}.
For undoped, nominally insulating $d^1$ materials, the criterion of being metallic is not 
always satisfied at low temperature. Uncorrelated $d^1$ materials are expected to be band metals. 
The samples we have studied, NbO$_2$ and LaTiO$_3$, are both  
metallic at high temperature and undergo a metal-to-insulator transition at 
1080 K (of Peierls type) \cite{0295-5075-58-6-851}  and 125 K (of Mott type) \cite{PhysRevLett.69.1796} respectively;
SrVO$_3$ is intrinsically metallic \cite{PhysRevLett.104.147601}.
Note that SrVO$_3$ already shows the $d^2$ peak in the optimally oxidized sample [Fig.~\ref{fig:SrVO3} (a)],
indicating its relatively strong charge fluctuation due to its metallic nature.

Specific to our experimental conditions, all $n$-doped SrTiO$_3$ are metallic at room temperature.
LaTiO$_3$ is already metallic at room temperature, 
which easily allows for charge fluctuation. For NbO$_2$, the sample is still nominally 
insulating at room temperature, but its relatively small band gap of $\sim$1.0 eV 
\cite{Posadas_NbO2.APL} likely results in non-negligible concentration of electrons 
in the conduction band at room temperature. The fact that no sample charging 
is observed during XPS measurements indicates that there is sufficient 
conductivity in the samples at room temperature (sufficient thermally excited carriers in the conduction band) 
to allow for charge fluctuation to occur. 
Therefore, although the charge fluctuation in the undoped, 
nominally insulating $d^1$ materials can be weaker compared to the doped samples, we believe it is still non-negligible.

\subsection{Relative importance of the initial-state and final-state effects}
We would like to conclude our theoretical analysis by addressing the relative importance of the initial-state and final-state effects.
From Eq.~\eqref{eqn:dp_cluster}, we see that the valence screening is described 
by the parameter $U_{dc}$, which is the strength of the core-hole-induced attractive potential. 
If $U_{dc} = 0$, then valence band electrons do not feel the existence of the core hole, and 
thus no final-state effect is involved.
With this observation, we  propose that the dimensionless parameter $\xi = U_{dc}/W$, with 
$W$ the typical energy scale of the valence bandwidth, can be 
used to characterize the relative importance between initial-state and final-state effects:
large $\xi$ favors the final-state effect; small $\xi$ favors the initial-state effect.
 As the bandwidth is proportional to the electron hopping $t$, we can roughly 
regard $1/U_{dc}$ as the time scale to create a core hole, and $1/t$ as the time scale  
for conduction electron to move to screen the core hole. Therefore the inverse of $\xi$ ($1/\xi$) essentially describes 
how efficient (fast) the conduction electrons screen the core hole.
By fixing the value of $U_{dc}$ (about 10 eV \cite{CoreLevel}), materials of large/small valence bandwidth 
favor the initial-state/final-state effect.

With this picture, we comment on the established interpretations of XPS spectra.
For covalent materials such as carbon and silicon, 
the initial-state appears to be dominant and 
the multi-peak structure is used appropriately  to signal the existence of different oxidation phases \cite{Miller_2002_C, PhysRevB.38.6084}.
Consistent with our argument, the diamond structure of C and Si 
indeed have relatively large valence bandwidths of approximately 20 eV \cite{PhysRevB.41.3048, Bassani} 
and 12 eV \cite{PhysRevB.10.5095},  respectively, which favors the initial-state effect.
For materials with valence electrons in localized orbitals (rare earth $4f$) such as 
lanthanum and cerium \cite{Kotani_1974, Kotani_99, PhysRevB.28.4315}, it is the final-state effect 
which dominates. For these materials the XPS multi-peak structure is not attributed to
the oxidation states, but can be used to determine material-specific model parameters by comparing to 
a model calculation \cite{CoreLevel}. A typical bandwidth of $f$-orbitals is about 4 eV 
\cite{PhysRevB.27.7330, PhysRevB.46.3458, PhysRevB.85.125134}, which favors the final-state effect.
In terms of valence bandwidth, the early transition metal oxides are in
between the two classes of materials  (about 6 to 8 eV \cite{Zollner,PhysRevB.79.113103}) .
%The ratio between $U_{dc}$ and maximum of $2D$, $\epsilon_p-\epsilon_d$.
As the experimental results for carefully grown samples from different probing techniques fit  the final-state effect better
(see also Refs.~\cite{PhysRevB.76.035101, PhysRevB.78.075103}), we believe 
the final-state effect is also the dominant one in the transition metal oxides.
Taking $U_{dc}$ to be 10 eV, we summarize the origin of the multi-peak structure in XPS for the materials mentioned
 above in Table \ref{table:xi}.

\begin{table}[h]
 \begin{tabular}{ l | l l l }
 Material & valence bandwidth ($W$) & $\xi = U_{dc}/W$ & origin of multiple peaks \\ \hline 
 diamond carbon & 20 eV & 0.5 & initial-state \\
 diamond silicon & 12 eV & 0.83 & initial-state \\
 SrTiO$_3$ & 6 eV & 1.67 & final-state \\
 CeNi$_2$ & 4 eV & 2.5 & final-state 
 \end{tabular}
 \caption{The origin of the multi-peak structure in XPS for various materials.
 The value SrTiO$_3$ is similar to the $d^1$ materials studied in this paper. 
 $U_{dc}$ is taken to be 10 eV.}
 \label{table:xi}
\end{table}

%%%%%%%%%%%%%%%%%%%%%%%%
\section{Conclusions}

We investigate the origin of the observed XPS multi-peak structure of single phase nominally $d^1$ transition metal oxides
including NbO$_2$, SrVO$_3$,  LaTiO$_3$, and lightly $n$-doped SrTiO$_3$.
Experimentally, we find that the XPS spectra (specifically the photoelectrons from Nb $3d$, V $2p$, Ti $2p$ core levels) 
of these materials all display at least two, and sometimes three pairs of peaks, 
which can be consistently assigned as $d^0$, $d^1$, and $d^2$ oxidation states.
For lightly $n$-doped SrTiO$_3$, a weak $d^1$ shoulder, 
whose energy position is independent of the dopants, appears with respect to the main $d^0$ peak.
For nominally $d^1$ transition metal oxides, 
electron doping increases the intensity of the $d^2$ peak but decreases that of the $d^0$ peak, whereas hole doping 
reverses this trend. A single $d^1$ peak is never observed, even in single phase samples.
In particular, the $d^0$ peak always exists even in the electron doped samples 
where stoichiometric analysis shows strong oxygen-deficiency and diffraction shows no secondary phases,
strongly indicating that the multi-peak structure is intrinsic to these materials.
Theoretically, we construct and solve a cluster-bath model, and explicitly demonstrate that 
the final-state effect (i.e. the valence response to the created core hole)
naturally leads to the multiple peaks in the XPS spectrum even in a spatially uniform system. 
Moreover, the relative peak strength as a function of doping is qualitatively consistent with 
the experimental observation. The combination of experimental and theoretical analysis 
leads us to conclude that the multi-peak structure in the nominally $d^1$ transition metal oxides 
is intrinsic, and does not necessarily imply the existence of spatially isolated (or clustered) $d^0$ and $d^2$ ions in a sample.
Using the same analysis, we argue that the ratio between the local screening potential and the valence bandwidth is the key 
dimensionless parameter that determines the relative importance between initial-state and final-state effects.
To establish the existence of different oxidation phases in a sample, further 
spatially-resolved probing techniques involving the valence electrons are needed. For this reason, 
investigating the final-state effect in x-ray absorption spectroscopy can be very helpful.

\section*{Acknowledgements}
%C.L. thanks Jeroen van den Brink and Nicholas Plumb for a few encouraging and enlightening conversations.
%We thank Andy O'Hara and Allan MacDonald for insightful comments.
C.L. thanks Jeroen van den Brink, Nicholas Plumb, and Ralph Claessen for encouraging and enlightening conversations.
We thank Miri Choi ((La,Sr)TiO$_3$), Daniel Groom (LaTiO$_3$) and Kristy Kormondy ((La,Sr)VO$_3$) for help in growth optimization,
and Andy O'Hara and Allan MacDonald  for insightful comments.
Support for this work was provided through Scientific Discovery through Advanced Computing (SciDAC) program 
funded by U.S. Department of Energy, Office of Science, Advanced Scientific Computing Research and 
Basic Energy Sciences under award number DESC0008877.

\appendix 
\section{Charge fluctuation in the metallic phase}
In Section IV we propose a model [Eq.~\eqref{eqn:bath_coupling}] and argue the importance 
of charge fluctuation  in the metallic phase. Here we use a very simple 
model to illustrate this effect. We consider a
three-site tight-binding model containing two electrons:
\beq 
H_{3o-TB} =-t  \sum_{ \sigma} (c^{\dagger}_{1,\sigma} c_{2,\sigma} + c^{\dagger}_{2,\sigma} c_{3,\sigma} 
+ c^{\dagger}_{3,\sigma} c_{1,\sigma} + h.c.),
\eeq
with $c_i$ the local orbital basis.
The two-particle ground state is $|\phi_0 \rangle = d^{\dagger}_{0,\uparrow} d^{\dagger}_{0,\downarrow} | vac \rangle$
with $d^{\dagger}_{0,\sigma} = (c^{\dagger}_{1,\sigma} + c^{\dagger}_{2,\sigma} + c^{\dagger}_{3,\sigma})/\sqrt{3}$.
Note that $d^{\dagger}_{0,\sigma}$ describes a Bloch orbital which is spatially extended.
When expressing the ground state using the local orbital basis, we have 
\beq 
|\phi_0 \rangle = \frac{1}{3} c^{\dagger}_{1,\uparrow} c^{\dagger}_{1,\downarrow} | vac \rangle +
\frac{1}{3} \left[ c^{\dagger}_{1,\uparrow} ( c^{\dagger}_{2,\downarrow} +  c^{\dagger}_{3,\downarrow} ) +
 ( c^{\dagger}_{2,\uparrow} +  c^{\dagger}_{3,\uparrow} ) c^{\dagger}_{1,\downarrow}
\right] | vac \rangle + 
\frac{1}{3} ( c^{\dagger}_{2,\uparrow} +  c^{\dagger}_{3,\uparrow} ) ( c^{\dagger}_{2,\downarrow} +  c^{\dagger}_{3,\downarrow} ) 
| vac \rangle.
\eeq
The terms are grouped according to the occupation on the first site.
In the local basis, we see that the many-body ground state (only two-body in this case) 
contains all doubly occupied, singly occupied, and unoccupied components; the situation
which is typically referred to as the charge fluctuation.
Note that this fluctuation has nothing to do with temperature, but  
originates solely from the many-body wave function.
The inclusion of the bath degree of freedom is to take this charge  fluctuation of the
local occupation into account.

\section{Details of computing core-level spectra}

\subsection{Basic formula}
The XPS spectrum is computed using \cite{CoreLevel, PhysRevB.28.4315}
\beq 
\begin{split}
\rho (\omega) &= \sum_n |\langle n (N-1) | c |GS \rangle|^2 \times \delta(\omega - [E_{GS} (N) - E_{core,n} (N-1) ] \,) \\
&= \frac{1}{\pi} \langle GS| c^{\dagger} \left[ \omega - (E_{GS} (N) - H_{tot}) - i \delta \right]^{-1} c|GS \rangle.
\end{split}
\label{eqn:XPS_rho_Appendix}
\eeq
In the calculation, we first solve the ground state $|GS \rangle$, and then follow the Lanczos procedure 
\cite{Grosso, RevModPhys.66.763} to compute the spectrum.

\subsection{Cluster model}
Here we compute $\rho (\omega)$ for the cluster model specified in Eq.(2) in the main text.
This problem can be exactly solved by diagonalizing only a $35 \times 35$ matrix.
The key numerical step, first realized by Gunnarsson and Sch\"onhammer in Ref. \cite{PhysRevB.28.4315},  
is that for all degenerate determinantal states, only one of their combinations contributes to
the exact ground state. In Table \ref{table:cubic_state} we list all 35 states.
The reference determinantal state, $|0 \rangle$, is defined by occupying all O $2p$ levels, and 
the other 34 states are labeled by particle-hole (p-h) pairs in $t_{2g}$ and $e_g$ sectors.

\begin{table}[ht]
\begin{tabular}{ l || l | l } 
 p-h pairs (notation) &label & ($n_t; n_{t2g}, n_{eg}$) [n]    \\ \hline
 0 ($|d^{0}\underline{L}^{0} \rangle$) & $|0 \rangle$ & (0; 0, 0)  [1]\\
 1 ($|d^{1}\underline{L}^{1} \rangle$) & $|1 \rangle$ to $|2 \rangle$ & (1; 1, 0), (1; 0, 1) [2]\\
 2 ($|d^{2}\underline{L}^{2} \rangle$)& $|3 \rangle$ to $|5 \rangle$ & (2; 2, 0), (2; 1, 1), (2; 0, 2)  [3]\\
 3 ($|d^{3}\underline{L}^{3} \rangle$)& $|6 \rangle$ to $|9 \rangle$ & (3; 3, 0), (3; 2, 1), (3; 1, 2), (3; 0, 3) [4]\\
 4 ($|d^{4}\underline{L}^{4} \rangle$)& $|10 \rangle$ to $|14 \rangle$ & (4; 4, 0), (4; 3, 1), (4; 2, 2), (4; 1, 3), (4; 0, 4)  [5]\\
 5 ($|d^{5}\underline{L}^{5} \rangle$)& $|15 \rangle$ to $|19 \rangle$ & (5; 5, 0), (5; 4, 1), (5; 3, 2), (5; 2, 3), (5; 1, 4) [5]\\
 6 ($|d^{6}\underline{L}^{6} \rangle$)& $|20 \rangle$ to $|24 \rangle$ & (6; 6, 0), (6; 5, 1), (6; 4, 2), (6; 3, 3), (6; 2, 4)  [5]\\
 7 ($|d^{7}\underline{L}^{7} \rangle$)& $|25 \rangle$ to $|28 \rangle$ & (7; 6, 1), (7; 5, 2), (7; 4, 3), (7; 3, 4) [4]\\
 8 ($|d^{8}\underline{L}^{8} \rangle$)& $|29 \rangle$ to $|31 \rangle$ & (8; 6, 2), (8; 5, 3), (8; 4, 4)  [3]\\
 9 ($|d^{9}\underline{L}^{9} \rangle$)& $|32 \rangle$ to $|33 \rangle$ & (9; 6, 3), (9; 5, 4) [2]\\
10 ($|d^{10}\underline{L}^{10} \rangle$) & $|34 \rangle$ & (10; 6, 4) [1]
\end{tabular}
\caption{States in the cluster with ten electrons.
$n_t$, $n_{t2g}$, $n_{eg}$ are the number of total p-h pairs, $t_{2g}$ p-h pairs, and $e_g$
p-h pairs respectively. [n] represents the number of states for the number of p-h pairs.
The diagonal energy is given by $U n_t (n_t-1)/2 - n_t \epsilon_p + n_{eg} \epsilon_{d} ( e_g) + n_{t2g} \epsilon_{d} ( t_{2g})$.}
\label{table:cubic_state}
\end{table}

\begin{figure}[http]
\epsfig{file=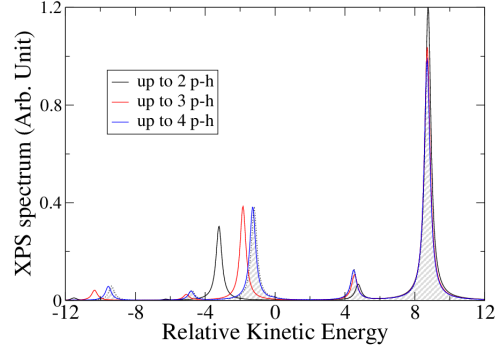, width = 0.4\textwidth}
\caption{ (Color Online) The p-h number dependence of the XPS spectra (arbitrary units) for the cluster model.
The shaded area represents the exact spectrum. We see that including states up to two p-h pairs already 
gives a very reasonable profile. Keeping states up to four p-h pairs almost 
reproduces the exact spectrum. 
}
\label{fig:XPS_p-h_dependence}
\end{figure}

To explicitly write down these states, we define the p-h operators as 
$P_i =  d^{\dagger}_i p_i $ for $i$ belongs to one of six $t_{2g}$ orbitals (including spins), 
and $\bar{P}_j =  d^{\dagger}_j p_j $ for $j$ belongs to one of four $e_{g}$ orbitals.
The state labeled as ($n_t; n_{t2g}, n_{eg}$) in Table \ref{table:cubic_state} is
\beq 
\begin{split}
(n_t; n_{t2g}, n_{eg}) \rightarrow \frac{1}{\sqrt{C^6_{n_{t2g}} } } \frac{1}{\sqrt{C^4_{n_{eg}} } } 
\left[ \sum_{\{i\}} \Pi_{ \{i\}} P_i \right] \left[ \sum_{\{j\}} \Pi_{ \{j\}} \bar{P}_j \right] | 0 \rangle.
\end{split}
\eeq
Here $\{i\}$ ($\{j\}$) represents all combinations of creating $n_{t2g}$ ($n_{eg}$) p-h pairs out of 
the reference state, and $ C^{N}_m \equiv \frac{N!}{(N-m)! m!}$.
Note that each individual determinantal state in the summation has the same energy, and it is
Gunnarsson and Sch\"onhammer's invaluable observation that only  the sum of them
contribute to the exact ground state, and all other combinations can be rigorously neglected. 
The coupling between these states is nonzero only when $|\Delta n_{eg}| + |\Delta n_{t2g}| = 1$, and 
can be computed straightforwardly. To check this formalism, we also computed the ground state and XPS spectrum
by diagonalizing the original $63504 \times 63504$ matrix (the dimension of 
filling 10 electrons in 20 orbitals with $M_z=0$), which gives identical 
results to that obtained by keeping only 35 states.
When including the bath degrees of freedom, it is not possible to include all states. 
In Fig.~\ref{fig:XPS_p-h_dependence} we show that keeping states up to two p-h pairs already 
results in a very reasonable profile. Keeping states up to four e-h pairs almost 
reproduces the exact spectrum.

\subsection{Cluster coupling to bath}
Now we solve the problem including both Eq.~\eqref{eqn:dp_cluster} and Eq.~\eqref{eqn:bath_coupling} in the main text.
There are six degenerate $t_{2g}$ and four degenerate $e_g$ d-p orbital pairs in the cluster, and  
each $3d$ orbital couples to its own bath. 
In the calculation, it is convenient to simply treat the O $2p$ orbital as one of the baths, whose
energy and coupling to the Ti $3d$ are respectively  $\epsilon_p (\Gamma)$ and $V (\Gamma)$.
In other words, we should notationally identify $b_{\epsilon_p,\Gamma} = p_{\Gamma}$,
and $V(\epsilon_p,\Gamma) = V (\Gamma)$.
The reference state $|  0 \rangle$ is chosen as
\beq
 | 0 \rangle \equiv \left[ \Pi_{i=1}^{10} \Pi_{\epsilon<\mu}  b^{\dagger}_{\epsilon, i} 
 p^{\dagger}_i \right]  |vac \rangle,
 \eeq
i.e. all bath levels below the chemical potential $\mu$ are filled.

\begin{figure}[http]
\epsfig{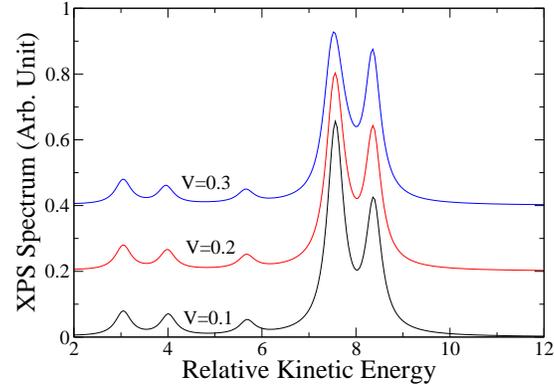}
\caption{ (Color Online) The computed XPS spectra for $\mu=1.0$ eV and V=0.1, 0.2, 0.3 eV.
The Ti$^{3+}$ peak emerges for all these $V$ values.
A Lorentz broadening of 0.2 eV is used. 
}
\label{fig:XPS_bath_gg_dependence}
\end{figure}

Similar to the previous subsection, we define the p-h operators for the $t_{2g}$ and $e_g$ sectors:  
%$P_i = d^{\dagger}_i p_i$, 
$P_i (\epsilon) = d^{\dagger}_i b_{\epsilon,i}$ 
$P_i (E,\epsilon) = b^{\dagger}_{E,i} b_{\epsilon,i}$ for $i \in t_{2g}$; 
 $\bar{P}_i (\epsilon) = d^{\dagger}_i b_{\epsilon,i}$ 
$\bar{P}_i (E,\epsilon) = b^{\dagger}_{E,i} b_{\epsilon,i}$ for $i \in e_{g}$. 
The $E$ and $\epsilon$ are for bath states which are higher and lower than the chemical potential.
We have tested energy spacing by discretizing the bath continuum into 32 to 100 intervals, 
and they result in essentially identical spectra.
We keep the states up to 2 p-h pairs, as tested in  Ref.~\cite{PhysRevB.28.4315}.
They are (in addition to the reference state)
\beq 
\begin{split}
 | \epsilon d \rangle_{t2g} &= \frac{1}{\sqrt{6}} \sum_{i=1}^6 P_i (\epsilon) |0\rangle, \\
 | \epsilon d \rangle_{eg} &= \frac{1}{\sqrt{ 4 }} \sum_{i=1}^{ 4 } \bar{P}_i (\epsilon)  |0\rangle, \\
 | \epsilon E \rangle_{t2g} &= \frac{1}{\sqrt{6}} \sum_{i=1}^6 P_i (E, \epsilon) |0\rangle, \\
 | \epsilon E \rangle_{eg} &= \frac{1}{\sqrt{ 4 }} \sum_{i=1}^{ 4 } \bar{P}_i (E,\epsilon)  |0\rangle, \\
 | \epsilon d,  \epsilon' d \rangle_{t2g} &= \frac{1}{\sqrt{6(6-1)}} \sum_{i \neq i'}
   P_i (\epsilon) P_{i'} (\epsilon') |0\rangle,\\
  | \epsilon d,  \epsilon d \rangle_{t2g} &= \frac{1}{\sqrt{6(6-1)/2}} \sum_{i<i'}
   P_i (\epsilon) P_{i'} (\epsilon) |0\rangle,\\
   | \epsilon d,  \epsilon' d \rangle_{eg} &= \frac{1}{\sqrt{4(4-1)}} \sum_{i \neq i'}
   \bar{P}_i (\epsilon) \bar{P}_{i'} (\epsilon') |0\rangle,\\
  | \epsilon d,  \epsilon d \rangle_{eg} &= \frac{1}{\sqrt{4(4-1)/2}} \sum_{i<i'}
   \bar{P}_i (\epsilon) \bar{P}_{i'} (\epsilon) |0\rangle,\\
 | \epsilon d,  \epsilon' d \rangle_{mix} &= \frac{1}{\sqrt{6(4)}} \sum_{i}^6 \sum_{i'}^{4}
   P_i (\epsilon) \bar{P}_{i'} (\epsilon') |0\rangle.
 \end{split}
\eeq
%To better describe the cluster, we further include nine more states, which are labeled as $|6\rangle$ to $|14\rangle$ 
%in Table \ref{table:cubic_state}.
The coupling between two states is non-zero only if the number of p-h pairs differs by one.
The XPS spectra are computed within these states. 
%We also show that the strength of bath-cluster coupling does not qualitatively affect 
%the final-state effect. 
Finally, in Fig.~\ref{fig:XPS_bath_gg_dependence}, 
we show the computed XPS spectra for $\mu=1$, and $V=$0.1, 0.2, 0.3 eV (defined in 
Eq.~\eqref{eqn:bath_coupling} in the main text), with
the Ti$^{3+}$ peak appearing for all of them.  
We emphasize again that the main role of the bath coupling is to introduce  
charge fluctuation within the cluster, and the form of the coupling 
plays a relative minor role in the spectrum.

\section*{Supplementary Materials}
\subsection*{RHEED data}

\begin{figure}[http]
\epsfig{file=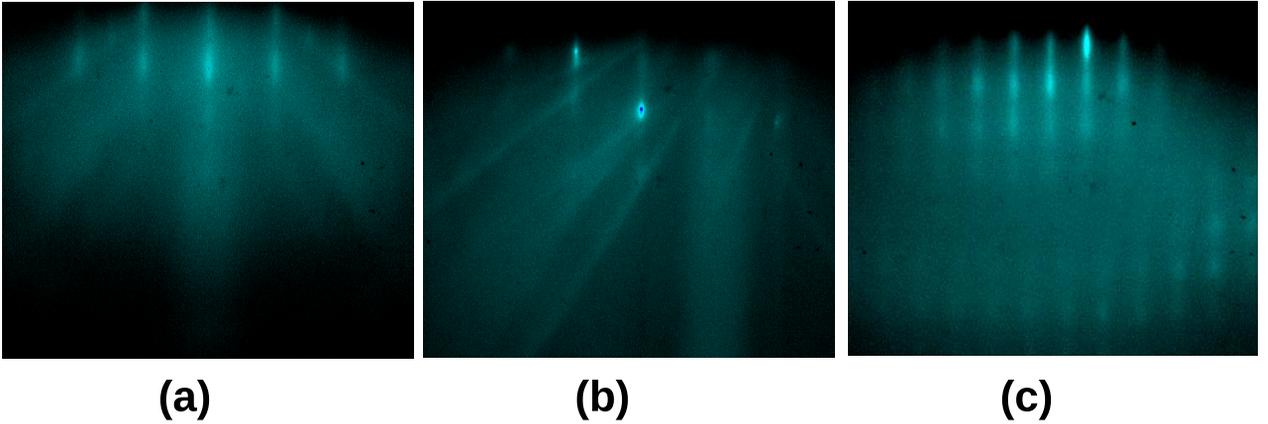, width = 0.95\textwidth}
\caption{ (Color Online) RHEED patterns for optimally oxidized epitaxial films of nominally $d^1$ transition metal oxides. 
(a) LaTiO$_3$ film grown on SrTiO$_3$ (100) viewed along the $\langle 110 \rangle$ azimuth. 
(b) SrVO$_3$ film grown on SrTiO$_3$ (100) viewed along the $\langle 110 \rangle$ azimuth. (c) NbO$_2$ film grown on SrTiO$_3$ 
(111) viewed along the $\langle 1-10 \rangle$ azimuth. For LaTiO$_3$ and SrVO$_3$ the RHEED pattern shows 
four-fold symmetry with a weak 2x reconstruction for LaTiO$_3$ and no reconstruction for SrVO$_3$. For NbO$_2$, 
the RHEED pattern shows six-fold symmetry due to the existence of three symmetry-related rotational domains.
}
\label{fig:RHEED}
\end{figure}

We provide in Fig. \ref{fig:RHEED}  the RHEED patterns for optimally oxidized epitaxial films of nominally 
$d^1$ transition metal oxides: (a) LaTiO$_3$; (b) SrVO$_3$; (c) NbO$_2$. 
For LaTiO$_3$ and SrVO$_3$ the RHEED pattern shows four-fold symmetry with a weak 2x reconstruction for LaTiO$_3$ 
and no reconstruction for SrVO$_3$. For NbO$_2$, 
the RHEED pattern shows six-fold symmetry due to the existence of three symmetry-related rotational domains.

\subsection*{Surface effect: SrVO$_3$ XPS spectra before and after Ar sputtering}

%\appendix

\begin{figure}[http]
\epsfig{file=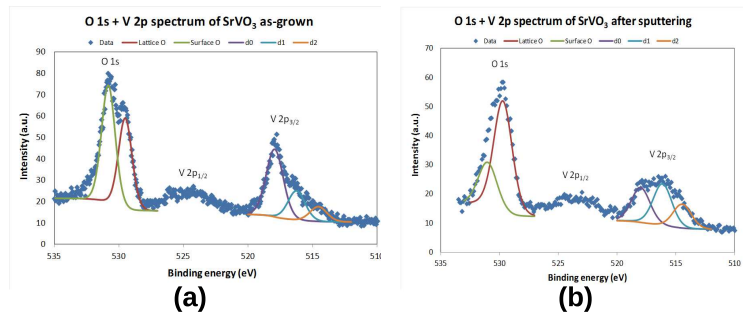, width = 0.6\textwidth}
\caption{ (Color Online) 
XPS spectrum of O $1s$ and V $2p$ region of SrVO$_3$ thin film before (a) and after (b) the Ar sputtering.
Note that even after the Ar sputtering removes surface atoms, the $d^{0}$ peak 
is still very significant. 
}
\label{fig:SrVO3_sputtering}
\end{figure}

Here we provide the SrVO$_3$ XPS spectra with and without Ar sputtering,
to show that {\it both} the surface reconstruction (initial-state effect) and final-state effect contribute to the multi-peak structure.
In situ XPS spectra of SrVO$_3$ thin films before and after sputtering
SrVO$_3$ is thin films exhibit a weak $\sqrt{2} \times \sqrt{2}$ surface reconstruction in low energy electron diffraction \cite{PhysRevB.80.235104}. 
This surface shows two components in the O $1s$ spectrum and at least two components in the V $2p$ spectrum. 
This is consistent with an oxygen-rich surface layer with V$^{5+}$ species present, 
even when the growth and XPS measurement are done in situ \cite{PhysRevB.80.235104, Takizawa_thesis}. 
Angle-dependent XPS indicates that this oxygen-rich 
layer is 2-4 \AA$ $ thick \cite{PhysRevB.80.235104}.
We performed in situ XPS measurements of a SrVO$_3$ thin film before and after argon ion sputtering of the surface to 
determine the effect of removal of the oxygen-rich surface layer on the O $1s$ and V $2p$ spectra of SrVO$_3$. 
A stoichiometric SrVO$_3$ film is grown by molecular beam epitaxy on a SrTiO$_3$(100) substrate at a temperature of
700$^{\circ}$C to a thickness of 150 \AA. At this thickness, none of the substrate photoelectrons is able to escape the surface, 
allowing for a clean determination of the oxidation states and stoichiometry of the film.  
XPS analysis is performed in a separate analysis chamber that is connected to the growth chamber by a vacuum transfer 
line with a base pressure of $1 \times 10^{-9}$ Torr. The analysis chamber has a base pressure of $3 \times 10^{-10}$ Torr. 
XPS spectra are taken before and immediately after sputtering. The sample is never exposed to the ambient at any time.
The surface sputtering is performed using argon ions generated by a differentially pumped Hiden IG 20 ion gun. 
The Ar background pressure is $5 \times 10^{-5}$ Torr using an electron emission current of 20 mA. 
The Ar ions are accelerated to an energy of 1 keV and rastered on 3 mm $\times$ 3 mm area of the sample resulting in a sample current of 120 nA. 
The ion sputtering is performed for a total time of 5 min.  Fig. \ref{fig:SrVO3_sputtering}(a) shows the O $1s$/V $2p$ spectrum before sputtering.

The spectrum before sputtering can be deconvoluted into two oxygen components (531.0 and 529.7 eV) and three vanadium components. 
The oxygen feature at 531.0 eV corresponds to strong surface oxygen feature consistent with the presence of an oxygen-rich surface layer. 
The oxygen feature at 529.7 eV corresponds to bulk oxygen in SrVO$_3$. 
The vanadium components have binding energies of 517.9, 516.2 and 514.5 eV corresponding to 3$d^0$ (V$^{5+}$), 3$d^1$ (V$^{4+}$), 
and 3$d^2$ (V$^{3+}$) components. The strong 3$d^0$ feature is consistent with a V$^{5+}$ surface layer 
as described by Takizawa et al. \cite{PhysRevB.80.235104}. 
Note that the oxygen-rich surface reconstruction is present even when the sample has not been exposed to excess oxygen.
The O $1s$/V $2p$ spectrum after sputtering is shown in Fig. \ref{fig:SrVO3_sputtering}(a). 
The same set of two oxygen components and three vanadium components are still present after removal of the surface atoms in the film. 
However, the surface oxygen component is reduced by a factor of 3/4 showing that most of the surface oxygen has been removed. 
The 3$d^0$ vanadium component is at the same time reduced by more than half (with a slight shift in binding energy to 518.1 eV). 
This shows that even after removal of the surface layer, the 3$d^0$ component is still present although significantly reduced in magnitude. 
The after sputtering spectrum is now more like the bulk-sensitive SrVO$_3$ spectrum reported by Eguchi et al. 
using hard x-ray photoemission \cite{Eguchi2007421}.

Because of the occurrence of an intrinsic surface reconstruction resulting in an oxygen-rich surface layer, 
which results in the presence of a 3$d^0$ feature in the V $2p$ spectrum, 
it is not possible to ascribe the observed 3$d^0$ feature as arising solely from the final state effect described in the main text. 
However, removal of the surface layer still shows the presence of a significant although strongly reduced 3$d^0$ component. 
While most of the 3$d^0$ signal in the as-grown SrVO$_3$ film is due to the surface reconstruction, there is a residual 3$d^0$ 
component in SrVO$_3$ present, which is what we ascribe to being due to the final state effect.

\bibliography{CoreLevelSpectrum2}

\begin{thebibliography}{67}
\bibitem[{bin()}]{binding}
\bibinfo{note}{This relation just states that electrons of smaller kinetic
  energy come from electronic states of higher binding energy.}
\bibitem[{not()}]{note1}
\bibinfo{note}{No peak asymmetry develops even when the photoelectron emission
  angle is varied from normal emission indicating no observable feature due to
  surface core level shifts. See also: Ref.~\cite{Heide}.}
\bibitem[{12_()}]{12_orbital}
\bibinfo{note}{The remaining 13 orbitals are linear combinations of O $2p$
  orbitals, with no Ti $3d$ component.}
\bibitem[{Clu()}]{Cluster_note}
\bibinfo{note}{As we are dealing with the case of low $d$ occupancy, the model
  keeps only the crystal field but neglects multiplet effects (Hund's
  coupling). The core-spin dynamics is also neglected for simplicity, as its
  main effect is to provide a spin-split peak [$2p_{3/2}$, $2p_{1/2}$ peaks
  Fig.~\ref{fig:XPS_STO}] \cite{core_spin_broadening}. The spin-orbit coupling
  of Ti, which is about only 0.03 eV, is also neglected
  \cite{PhysRevB.87.161102}.}
\bibitem[{Fit()}]{Fitting}
\bibinfo{note}{In particular, the hybridization parameters are from ARPES, the
  crystal field and energies of Ti $3d$, O $2p$ are from ellisometry data, and
  $U_{dc}$, $U$ are derived from XPS. We note that except for $U$ and $U_{dc}$,
  all other parameters are fixed by the band structure and are robust. $U_{dc}$
  is chosen to fit the experimental satellite peak position, and gives similar
  profile within the range 7-10 eV. The value of $U$ turns out not to affect
  the XPS spectrum much when it is larger than 5 eV.}
\bibitem[{key()}]{key_trick}
\bibinfo{note}{The invaluable observation by Gunnarsson and Sch\"onhammer can
  be summarized by the following statement: among all degenerate determinantal
  states, only one of their combinations contributes to the exact many-body
  ground state.}
\end{thebibliography}

%\subsection{Cluster model}

%\appendix

%\section{Experimental details}

\end{document}